\documentclass{osa-article}
\journal{osajournal}

\articletype{Research Article}
\usepackage{subfig}
\usepackage{caption}
\usepackage{bm}
\begin{document}
\title{Near-resonant light transmission in two-dimensional dense cold atomic media with short-range positional correlations}

\author{B. X. Wang, C. Y. Zhao\authormark{*}}

\address{Institute of Engineering Thermophysics, School of Mechanical Engineering, Shanghai Jiao Tong University, Shanghai 200240, China}
\email{\authormark{*} changying.zhao@sjtu.edu.cn} 


\begin{abstract}
Light propagation in disordered media is a fundamental and important problem in optics and photonics. In particular, engineering light-matter interaction in disordered cold atomic ensembles is one of the central topics in modern quantum and atomic optics. The collective response of dense atomic gases under light excitation, which crucially depends on the spatial distribution of atoms and the geometry of the ensemble, has important impacts on quantum technologies like quantum sensors, atomic clocks and quantum information storage. Here we analyze near-resonant light transmission in two-dimensional dense ultracold atomic ensembles with short-range positional correlations. Based on the coupled-dipole simulations under different atom number densities and correlation lengths, we show that the collective effects are strongly influenced by those positional correlations, manifested as significant shifts and broadening or narrowing of transmission resonance lines. The results show that mean-field theories like Lorentz-Lorenz relation are not capable of describing such collective effects. We further investigate the statistical distribution of eigenstates, which are significantly affected by the interplay between dipole-dipole interactions and position correlations. This work can provide profound implications on collective and cooperative effects in cold atomic ensembles as well as the study of mesoscopic physics concerning light transport in strongly scattering disordered media.
\end{abstract}


\section{Introduction}\label{intro}
Light propagation in disordered media is a fundamental and important problem in many fields of science and engineering, such as optics and photonics \cite{sheng2006introduction,Rotter2017}, radiative heat transfer \cite{tien1987thermal,modest2013radiative}, astrophysics and remote sensing \cite{tsangRS2000,mishchenko2006multiple}, soft mater physics and chemistry \cite{vanmegenPRA1991,manPRL2013}, biomedical engineering \cite{horstmeyer2015guidestar} and so on. In optics and photonics, harnessing the interaction between light and disordered photonic structures recently inspires the rapid and amazing development of the field of "disordered photonics" \cite{wiersma2013disordered,lopezADOM2018}, in which focusing \cite{Vellekoop2007,Vellekoop2010} and imaging \cite{Mosk2012} through some seemingly scrambled media were successfully demonstrated. Moreover, enormous applications can be achieved by using disordered media with engineered micro/nano-structures, such as spontaneous emission control \cite{garciaAnnPhys2017}, random lasers \cite{Cao1999,Wiersma2008,linACSPhoton2017} and radiative cooling \cite{zhaiScience2017,baoSEMSC2017,liScience2019}, etc.

As disorder in micro/nano-structures leads to complicated multiple scattering and interference phenomena of light, consequently, a plethora of interesting transport behaviors like coherent backscattering and Anderson localization \cite{wiersma1997localization,Segev2013}, position-dependent diffusion constant \cite{kopPRL1997}, sub-diffusion \cite{sebbahPRB1993} and super-diffusion \cite{barthelemyNature2008,bertolottiPRL2010} can emerge. Moreover, engineering the disorder via structural correlations in the micro/nano-structures allows people to manipulate light scattering and interference in disordered materials \cite{conleyPRL2014}. Structural correlations are the reminiscence of order (usually short- or medium-ranged) in the spatial variation of dielectric properties in disordered media \cite{lopezADOM2018}, which can give rise to definite phase differences among scattered waves \cite{laxRMP1951,laxPR1952,tsang2004scattering2} and hence affect the transport properties of light significantly. For instance, the simplest hard-sphere-type correlation which exists in densely packed hard spheres \cite{wertheimPRL1963} can usually increase the transport mean free path of light \cite{fradenPRL1990,reuferAPL2007}. Screened Coulomb potential induced positional correlations in nanoparticle colloidal suspensions were found to induce strongly negative asymmetry factor \cite{rojasochoaPRL2004,wangJAP2018}. Recently, some unexpected optical properties like optical transparency \cite{leseurOptica2016}, enhanced absorption \cite{liuJOSAB2018} and isotropic photonic band gaps \cite{Florescu2009,Froufe-Perez2016,Froufe-PerezPNAS2017,ricouvierPNAS2019} were discovered in 2D and 3D disordered media with certain structural correlations, e.g., the stealthy hyperuniform media. Therefore, structural correlations offer a great opportunity for tailoring the optical properties of disordered materials, with promising applications like structural coloration \cite{xiaoSciAdv2017,hwang2019effects}, bright white paints \cite{pattelliOptica2018} and solar energy harvesting \cite{Vynck2012,Fang2015JQSRT,Liew2016ACSPh,liuJOSAB2018,bigourdanOE2019}, etc. 


Despite these fascinating advancements, however, it is still not fully understood how structural correlations affect optical properties of disordered media, even for an ideally simple system composed of randomly distributed spherical scatterers \cite{Naraghi2015,Naraghi2016,wangPRA2018,pattelliOptica2018,ramezanpourJQSRT2019}. In particular, the interplay between structural correlations and near-field as well as far-field electromagnetic interactions among scatterers is still not very clear \cite{Naraghi2015,Naraghi2016,escalanteADP2017}, especially near single scatterer internal resonances (like Mie resonances for dielectric nanoparticles) at high packing density  \cite{lagendijk1996resonant,sapienzaPRL2007,garciaPRA2008,aubryPRA2017,tallonPRL2017}.

On the other hand, the last three decades have witnessed the rapid development of laser cooling and trapping of atoms to realize ultracold atomic gases with extremely low temperatures on the order of a few nanokelvin \cite{cohentannoudji2011}, which stimulate a wide range of exciting applications like high-precision atomic clocks, quantum information processing, quantum computing and quantum simulation of condensed matter systems and so on \cite{cohentannoudji2011}. Regarding the very high resonant scattering cross section  of a single two-level cold atom at the dipole transition (on the order of $\sim\lambda_0^2$, where $\lambda_0$ is the wavelength of transition, thus surprisingly larger than the size of a single atom), cold atomic systems are capable of achieving strong light-matter interaction \cite{cohentannoudji2011}. In particular, disordered cold atomic gases offer new opportunities for theoretical and experimental study of multiple scattering of light, and they are advantageous over conventional discrete random media due to the precisely controllable and highly tunable system parameters in an unprecedented regime \cite{baudouin2014cold,haveyPhysrep2017}. 
Multiple scattering phenomena, for instance, coherent backscattering \cite{labeyrieJOB2000}, slow diffusion \cite{labeyriePRL2003}, Anderson localization \cite{Skipetrov2015,moreira2019localization,maximo2019anderson} and random lasing \cite{baudouinNaturephys2013} were investigated, along with some interesting phenomena like L\'evy flights of photons \cite{pereiraPRL2004,mercadierPRA2013}, thermal decoherence \cite{labeyriePRL2006} and non-Lorentzian transmission spectra \cite{zhuPRA2016,cormanPRA2017}, etc. Collective and cooperative effects that are difficult to observe in conventional condensed matter systems also emerge in cold atomic systems, such as the collective polaritonic modes \cite{schilderPRA2016}, superradiant and subradiant collective modes \cite{guerin2017light}, collective Lamb shift \cite{keaveneyPRL2012,meirPRL2014} and so on. Moreover, the easy incorporation of nonlinearity can provide a platform for the study of multiple scattering of interacting photons, where more intriguing many-body physics can take place \cite{changNaturephoton2014}, not to mention that the quantum nature of cold atoms (for instance quantum statistical correlations in Fermi-Dirac gases \cite{ruostekoskiPRL1999,ruostekoskiPRA2000}) might also affect light propagation, which very much enriches the underlying physics.

Structural correlations indeed exist and can also be engineered in cold atomic gases. It is commonly assumed in many theoretical treatments that the positions of atoms are completely independent of each other, especially in very dilute gases, e.g., Refs. \cite{chomazNJP2012,pellegrinoPRL2014,Skipetrov2014,Skipetrov2015,jenneweinPRL2016,schilderPRA2016,zhuPRA2016,ruostekoskiOE2016,Cherroret2016,javanainenPRA2017,cormanPRA2017,skipetrov2019transport}. Strictly speaking this assumption is not true because at very short distances atoms can interact with each other through, like, van der Waals interactions and collisions (like $s$-wave scattering) \cite{grossPhysRep1982,cohentannoudji2011}. Quantum statistics at low temperature can lead to significant correlations of atom positions in both trapped Bose and Fermi gases \cite{moricePRA1995,parkinsPhysrep1998,ruostekoskiPRL1999,ruostekoskiPRA2000}, where  the large de Broglie wavelength of atoms introduces a significant correlation length. For instance, low-density Fermi gas at zero temperature exhibits a positional correlation length on the order of $k_F^{-1}$, where $k_F\propto(6\pi^2\rho)^{1/3}$ is the Fermi wave number and $\rho$ is atom number density \cite{ruostekoskiPRL1999}. These positional correlations have nontrivial consequences on the optical interactions in atomic gases \cite{moricePRA1995,parkinsPhysrep1998,demacroPRA1998,ruostekoskiPRL1999,ruostekoskiPRA2000}. For instance, Ruosteskoski and Javanainen \cite{ruostekoskiPRL1999} theoretically showed that a dramatic line narrowing can occur in the Fermi gas even at low densities. Recently, Peyrot \textit{et al.} \cite{peyrotPRL2018} also implied that short-range interactions induced correlations as well as light-induced correlations can be a possible source of experimentally measured nonlocality in the optical response in high density atomic ensembles. The negligence of positional correlations may be a possible reason for the deviations between conventional coupled dipole simulations assuming fully disordered atoms and some experimental measurements \cite{pellegrinoPRL2014,jenneweinPRL2016,cormanPRA2017} besides other explanations like residual motion of the atoms during the probing, nonlinear effects and complex atomic level structures, etc. In addition, another method of creating positional correlations is to directly synthesize the spatial profile of atoms. Recent developments in the manipulation of single atoms make the fabrication of atomic lattices with \textit{arbitrary} distributions feasible. It was already shown that 1D $\mathrm{^{87}Rb}$ atomic arrays with desired arrangements in a large scale (more than 50 atoms) can be assembled in an atom by atom way, through a fast, real-time control of an array of tightly focused optical tweezers \cite{endresScience2016}. Two-dimensional $\mathrm{^{87}Rb}$ arrays with user-defined geometries can be also fabricated in a similar way \cite{lahayeScience2016}. Other modern quantum simulation techniques, such as nanophotonic atom lattices using dielectric photonic crystals \cite{gonzalezNaturephoton2015} and plasmonic nanoparticle arrays \cite{gullansPRL2012}, are also possible ways. These techniques offer great opportunities for the study of multiple scattering of light in controlled structural correlations at the most fundamental level.

In the present work, we study near-resonant light transmission in a correlated two-dimensional cold atomic gas, in order to acquire a deep understanding of the effect of positional correlations on resonant multiple scattering in disordered media. The positional correlations among atoms are introduced by adding an exclusion volume for each atom and implementing a Metropolis sampling process, which is manifested as random motions and collisions of the atoms in the 2D plane and can make atom positions strongly correlated. Based on the coupled-dipole simulations under different atom number densities and correlation lengths, we show that the collective effects are strongly influenced by those positional correlations, which lead to significant shifts and broadening or narrowing of transmission resonance lines. We further investigate the distribution and statistics of eigenstates, which are significantly affected by the atom number density and correlation length. The results show that mean-field theories like Lorentz-Lorenz relation are not capable of describing such collective effects. This work may provide profound implications on collective and cooperative effects in cold atomic ensembles, which are crucial for modern quantum technologies, as well as the study of mesoscopic physics concerning light transport in strongly scattering disordered media.


\section{Methodology}
\begin{figure}[htbp]
	\centering
	\subfloat{	
		\label{fig1a}
		\includegraphics[width=1\linewidth]{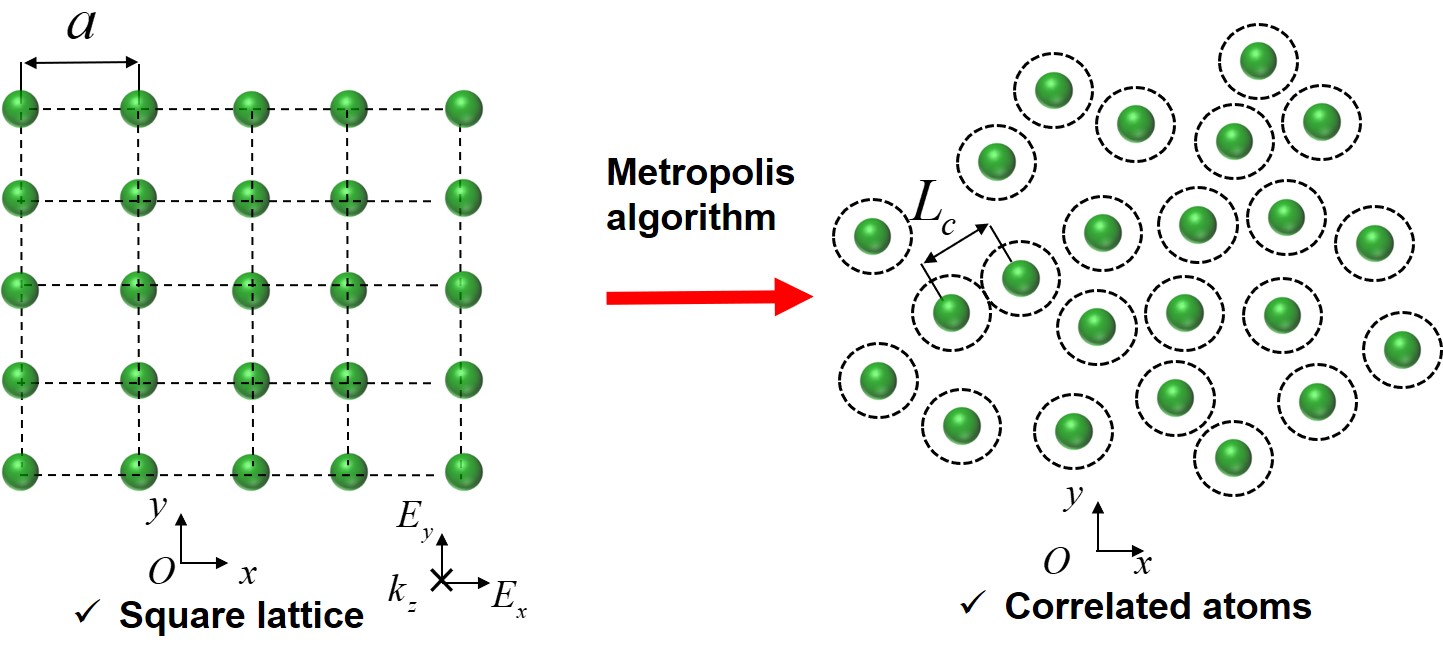}\label{atomgas}
		
	}
	\caption{Schematic of the 2D disordered atomic gas located in $xOy$ plane at $z=0$. The disordered positions of atoms (right) are generated through a Metropolis algorithm \cite{tsang2004scattering2}, starting from a square lattice with a period of $a$ (left). The normally incident light propagates along $z$-axis with a wavenumber $k_z$.}\label{config}
\end{figure}

Here we study a 2D cold atomic gas expanding in the $xOy$ plane, as shown in Fig. \ref{atomgas}. Because the vanishingly small transmission of high-density 3D samples introduces difficulties to experimental detection \cite{cormanPRA2017}, 2D and quasi-2D atomic gases provide a desirable platform for studying the resonant dipole-dipole interactions in dense cold atomic ensembles in a microscopic level. Trapped (quasi-) 2D atomic gases can be prepared, for example, using single pancake-like harmonic traps \cite{gorlitzPRL2001} or at the nodes of one-dimensional (1D) optical lattice potentials \cite{burgerEPL2002}. 2D gases in harmonic trapping potentials are usually inhomogeneous, and homogeneous 2D Bose \cite{chomazNaturecomms2015} and Fermi \cite{hueckPRL2018} gases trapped in box potentials were also reported recently. The cold atoms are assumed to be frozen in a disordered distribution of positions during the radiation transfer process, i.e., under a quenched disorder. This assumption is valid for cold atoms whose thermal velocity satisfies $kv\ll\gamma$. In other words, the Doppler broadening $\Delta_D$ much smaller than the transition linewidth $\Gamma$ in the low excitation limit \cite{zhuPRA2016,bromleyNComms2016}, thus, leading to a homogeneously broadened gas. Light pressure and effects of photon recoils of atoms are also neglected, which is valid when $2mk_BT\gg(\hbar k)^2$, where $m$ is the mass of atom.
For simplicity, we only consider two-level atoms, which have three degenerate excited states 
denoted by $|e_{\alpha}\rangle$ polarized along different directions, where $\alpha=x,y,z$ stands for the Cartesian coordinates, with a ground state denoted by $|g\rangle$. Under the low excitation limit, we can apply the single excitation approximation and work in the subspace spanned by the ground states $|G\rangle\equiv|g...g\rangle$ and the single excited states $|i\rangle\equiv|g...e_i...g\rangle$ of the atoms \cite{guerinPRL2016}. Moreover, we can adiabatically eliminate the photonic degrees of freedom in the reservoir (i.e., the quantized electromagnetic field).
After these assumptions, light-matter interaction can be described by classical electrodynamics under the following coupled-dipole equation as \cite{guerinPRL2016,wangOE2017}
\begin{equation}\label{coupled-dipole}
\mathbf{p}_j(\omega)=\alpha(\omega)\left[\mathbf{E}_\mathrm{inc}(\mathbf{r}_j)+\frac{\omega^2}{c^2}\sum_{i=1,i\neq j}^{N}\mathbf{G}_0(\omega,\mathbf{r}_j,\mathbf{r}_i)\mathbf{p}_i(\omega)\right],
\end{equation}
where $\mathbf{p}_j(\omega)$ is the excited dipole moment of $j$-th atom, $\mathbf{E}_\mathrm{inc}$ is the electric field of the incident light. Here $\alpha(\omega)$ is the polarizability of the two-level atom, which under our assumptions is described as \cite{ruostekoskiPRL2016,schilderPRA2016,wangOE2017}
\begin{equation}
\alpha(\omega)=-\frac{6\pi c^3}{\omega^3}\frac{\gamma/2}{\omega-\omega_0 + i \gamma/2},
\end{equation}
where the transition angular frequency of the two-level atom is ${\omega_{0}}$ and transition linewidth of $\gamma$ with $\omega_{0}\gg\gamma$.  $\mathbf{G}_0(\omega,\mathbf{r}_j,\mathbf{r}_i)$ is the free-space dyadic Green's function in the following form
\begin{eqnarray}
\mathbf{G}_{0}(\omega,\mathbf{r}_j,\mathbf{r}_i)=\frac{\exp{(ikr)}}{4\pi r}\Big[\left(\frac{i}{kr}-\frac{1}{k^2r^2}+1\right)\mathbf{I}+\left(-\frac{3i}{kr}+\frac{3}{k^2r^2}-1\right)\mathbf{\hat{r}}\mathbf{\hat{r}}\Big],
\end{eqnarray}
where the singular part (associated with the Dirac delta function) of Green's tensor is not explicitly expressed because the circumstances that atoms overlap are excluded, and $r=|\mathbf{r}|=|\mathbf{r}_j-\mathbf{r}_i|>0$ is the distance between two atoms, $\mathbf{\hat{r}}$ is the unit vector with respect to $\mathbf{r}$. Moreover, here we consider that a linearly polarized plane wave is normally illuminated to the atomic sample, which is described as $\mathbf{E}_{\rm{inc}}(\mathbf{r})=\mathbf{E}_0\exp{(i\mathbf{k}\cdot\mathbf{r})}$ with a wavevector $\mathbf{k}=k\hat{\mathbf{e}}_z$, where $\hat{\mathbf{e}}_z$ is the unit vector along $z$-axis. The electric field of incident light is thus polarized in the $xOy$ plane. 

Direct numerical calculation can be implemented to obtain the dipole moments of atoms given the positions of atoms, and the total transmission coefficient for a finite lattice can be computed by following the method in Ref. \cite{chomazNJP2012}, in which a lens is put in the far-field downstream of the atomic ensemble, whose axis is along $+z$ direction and located at $z=z_L\gg2\pi/k$ with a screen loclated in the focus of the lens. The total transmission coefficient is obtained by collecting all light transmitted to the lens and focused to the screen using the diffraction theory as \cite{chomazNJP2012,yooOE2016,wangOE2017}
\begin{equation}\label{eq_trans}
t_\beta=1+\frac{ik}{2E_0L_xL_y}\sum_{i=1}^{N}p_{i,\beta}\exp(-ikz_i),
\end{equation} 
where $L_x$ and $L_y$ are the boundary lengths of the simulated finite 2D ensemble, and $\beta$ denotes $x,y,z$. To avoid side effects, in simulation $kL_x\gg kz_L\gg1$ and $kL_y\gg kz_L\gg1$ should be guaranteed. Since the 2D arrays are located at $z_i=0$ for each atom, Eq.(\ref{eq_trans}) can be written in a more compact fashion as $t_\beta=1+ik\sum_{i=1}^{N}p_{i,\beta}/(2E_0L_xL_y)$.
Total transmittance defined for light intensity is then calculated as 
\begin{equation}
T_\beta=|t_\beta|^2.
\end{equation}

In this paper we focus on one of the simplest types of positional correlations, which is introduced by adding an exclusion volume for each atom and implementing a Metropolis sampling process starting from a square lattice of atoms with a lattice constant of $a$ \cite{chomazNJP2012}. This process is manifested as random motions and collisions of the atoms in the 2D plane and can make atom positions strongly correlated. The exclusion volumes of atoms result in a correlation length of $L_c$, which is the minimum possible distance between different atoms. Here we choose several cases for $L_c$, ranging from 0 (fully disordered) to $0.8a$ (strongly correlated).
The methodology used in this paper, i.e., synthetic positions of atoms and coupled-dipole model are successfully used to interpret experimental results of collective effects in dilute as well as dense cold atomic samples \cite{javanainenPRA1999,chomazNJP2012,ruostekoskiPRL2016,javanainenPRA2017}, which can be bosonic \cite{ruostekoskiPRL2016} or fermionic \cite{ruostekoskiPRL1999} provided the quantum statistics is accurate. Therefore, we believe this methodology can provide adequate physical details of light-matter interaction, especially the resonant dipole-dipole interactions, in this many-body system.

\section{The case of a dilute gas}

\subsection{Transmission and optical depth}
\begin{figure}[htbp]
	\centering
	\subfloat{
		\includegraphics[width=0.46\linewidth]{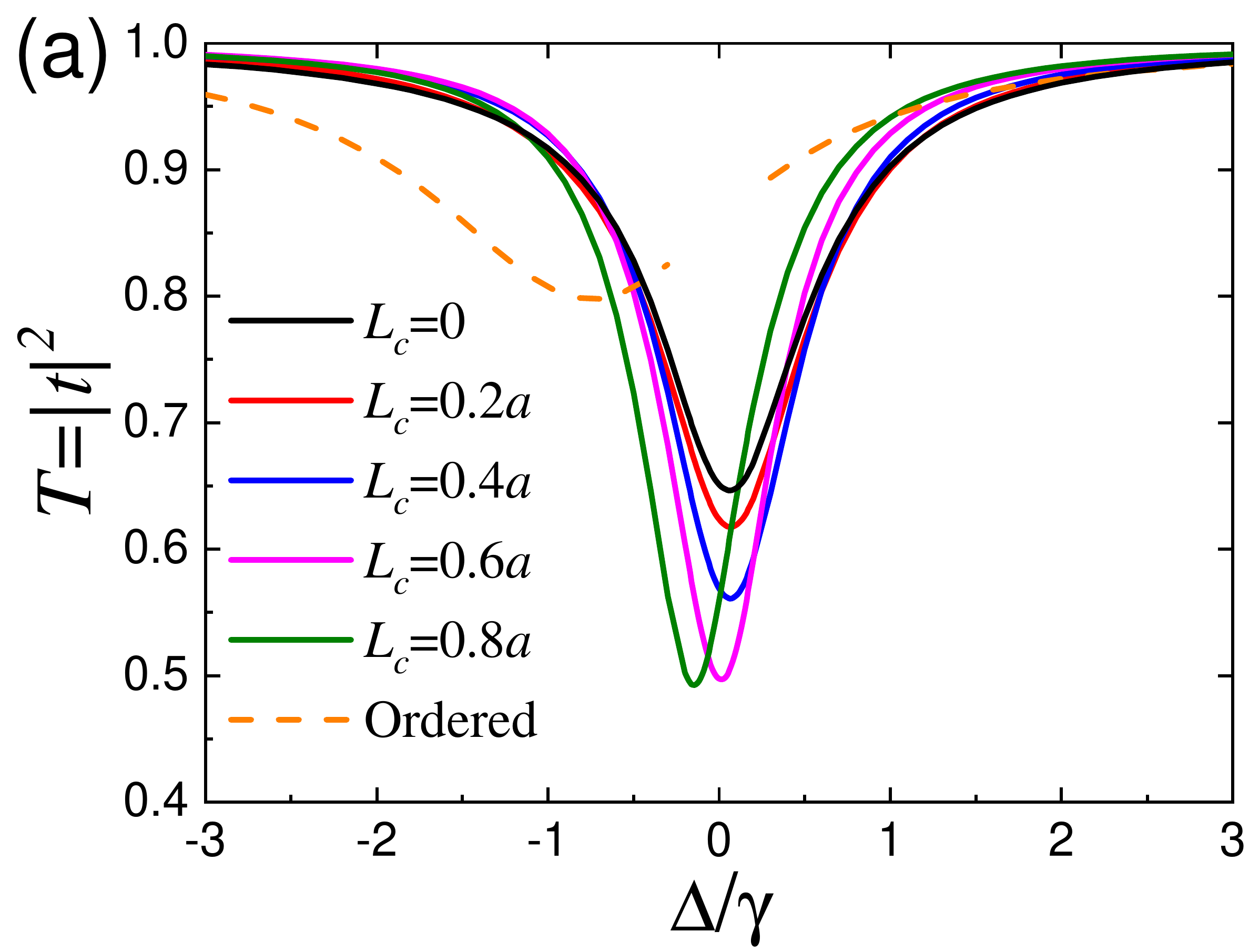}\label{transmittance_ratio1}
	}
	\hspace{0.01in}
	\subfloat{
		\includegraphics[width=0.46\linewidth]{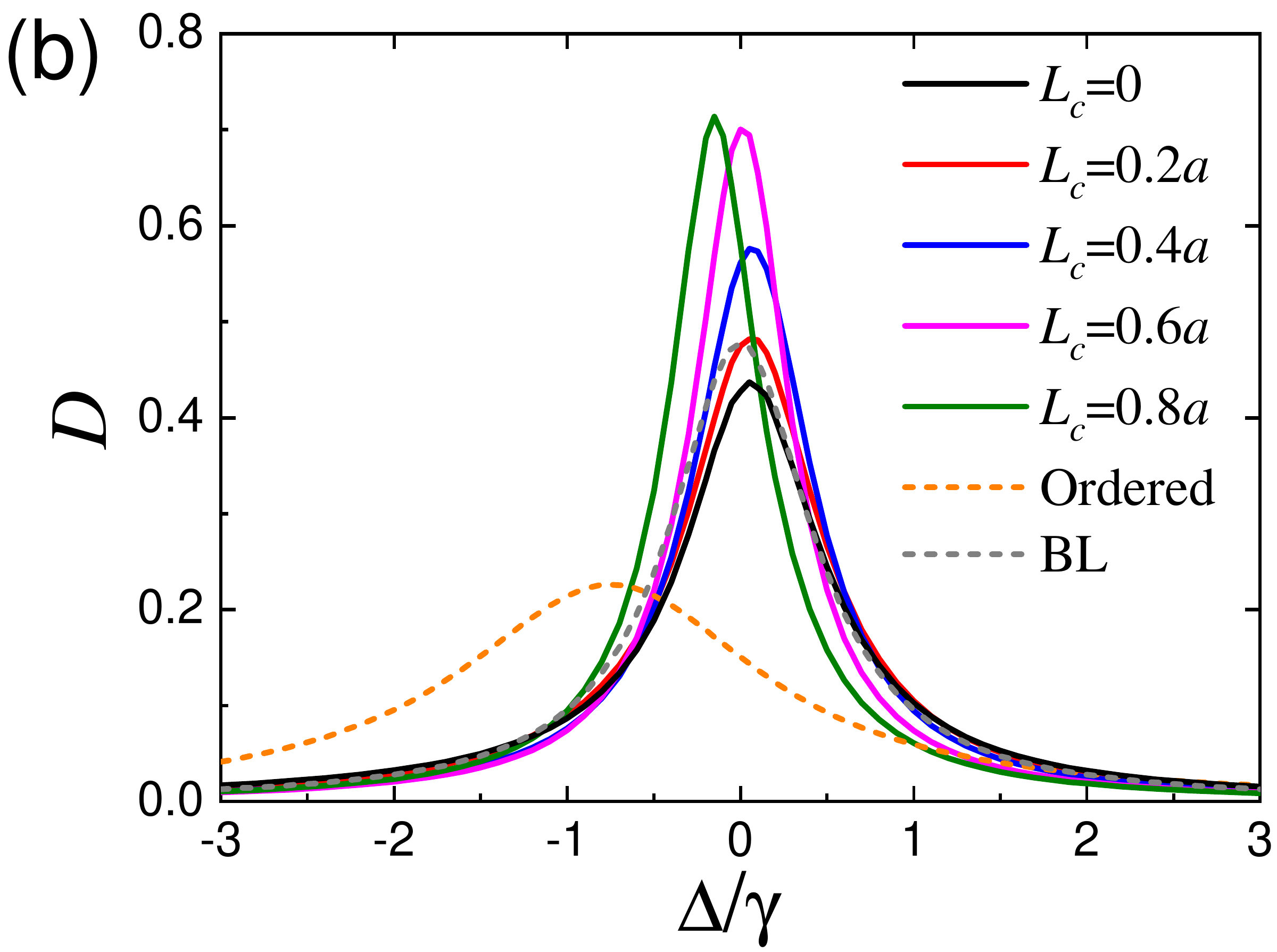}\label{OD_ratio1}
	}
	\caption{Transmittance spectra for 2D atomic arrays with different correlation lengths $L_c$, compared with the ordered square lattice under $a/\lambda_0=1$. (a) Total transmission $T$. (b) Optical depth $D$.}\label{transmittance_a1}
\end{figure}

In this subsection, we discuss these collective effects on the transmission spectra. We first study the case of $a/\lambda_0=1$ which stands for an atom number density of $n_0=1/\lambda_0^2$ in the initial square lattice, which is relatively dilute compared to some dense atomic samples. The transmission spectra for different correlation lengths are shown in Fig.\ref{transmittance_ratio1}, compared with the ordered square lattice with a lattice constant $a/\lambda_0=1$. 

It is observed from Fig.\ref{transmittance_ratio1} that the positional correlations influence the near-resonant responses of disordered atomic samples significantly, while at large detunings, these spectra approach each other. This is because the dipole-dipole interactions are strong near the single atom resonance and very sensitive to the distances between different atoms, which are associated with the correlation lengths among them. At large detunings, the dipole-dipole interactions become negligible and the single atom scattering behavior recovers, leads to similar responses which only depend on the number density. In particular, the valley (minimum) of the transmission spectra of disordered atomic samples decreases with the increase of the correlation length. This phenomenon can be understood by the role of recurrent scattering effect, which usually lowers the overall extinction coefficient and is significant when the interatomic distance is small \cite{Cherroret2016}. When the correlation length increases, the recurrent scattering effect is reduced, which leads to higher total extinction and thus smaller transmission. 

On the other hand, the spectral position of the transmission valley slightly redshifts from weakly blue-detuned to weakly red-detuned. This frequency shift cannot be explained through the local-field shift, or the Lorentz-Lorenz (LL) shift, since the LL shift only depends on the atom number density. Moreover, LL relation is only valid when the atoms are totally uncorrelated in both their positions and optical responses \cite{javanainenPRL2014}. 
These simulation results suggest that since the sign of collective frequency shift sensitively depends on the positional correlations, which can provide a possible explanation for recent confusing observations of blue shift \cite{cormanPRA2017} and red shift \cite{jenkinsPRL2016} in dense cold atomic gases.

In addition, the transmission spectra of disordered atom samples are vastly different from that of the ordered lattice, which exhibits an overall high transmission near the valley at $\Delta/\gamma\sim-0.7$, which results from the long-range order of the lattice. This difference explicitly shows the distinction between short-range order and long-range order, and in the lattice case the optical response is very much governed by the Bloch modes extended over the entire system.

Furthermore, in order to demonstrate the interference nature of present results, we also show the transmittance spectrum calculated by independent scattering approximation (ISA) as  
\begin{equation}
T_\mathrm{ind}(\omega)=\exp{\left[-n_0\sigma(\omega)\right]},
\end{equation} 
where $\sigma(\omega)=6\pi/[k^2(4\Delta^2/\gamma^2+1)]$ is the single scattering cross section of individual atoms. ISA, as a zeroth order approximation for multiple scattering problem\cite{vanRossumRMP1999}, treats all atoms scatter light independently without any considerations on  mutual interference effects of scattered waves. In this condition, the optical depth (OD) follows the Beer's law (BL) as
\begin{equation}
D_\mathrm{BL}=-\ln{T_\mathrm{ind}(\omega)}=n_0\sigma(\omega).
\end{equation}
The results of numerically calculated OD and OD under BL are also presented in Fig.\ref{OD_ratio1}. It is seen that the result of BL deviates significantly from the all the numerically simulated cases under different correlation lengths at small detunings, while at large detunings, all the results approach each other due to the recovery of single scattering behavior.


\subsection{Eigenstate analysis}
\begin{figure}[htbp]
	\centering
	\subfloat{
		\includegraphics[width=0.25\linewidth]{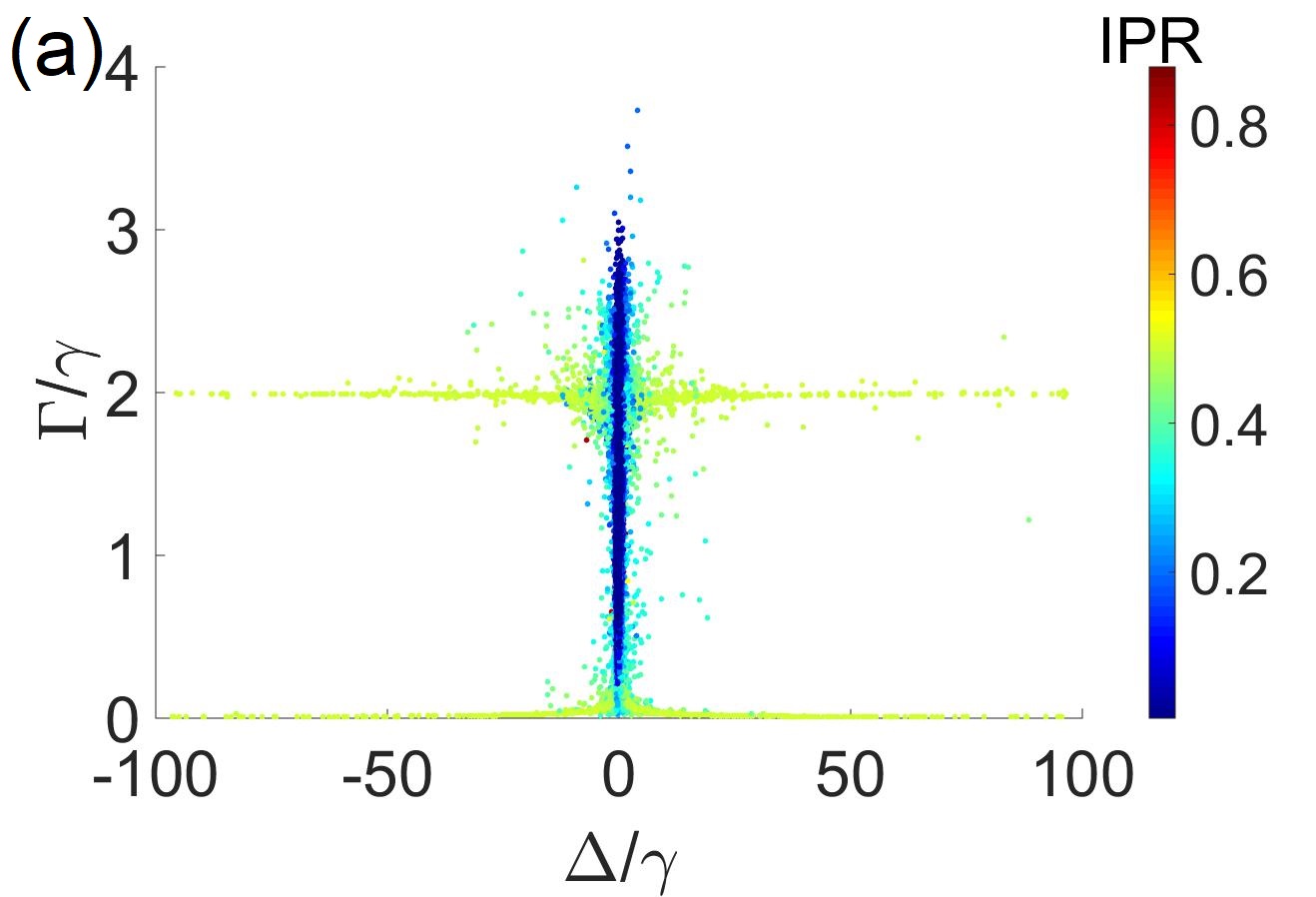}\label{eigenmodecor0}
	}
	\hspace{0.01in}
	\subfloat{
		\includegraphics[width=0.25\linewidth]{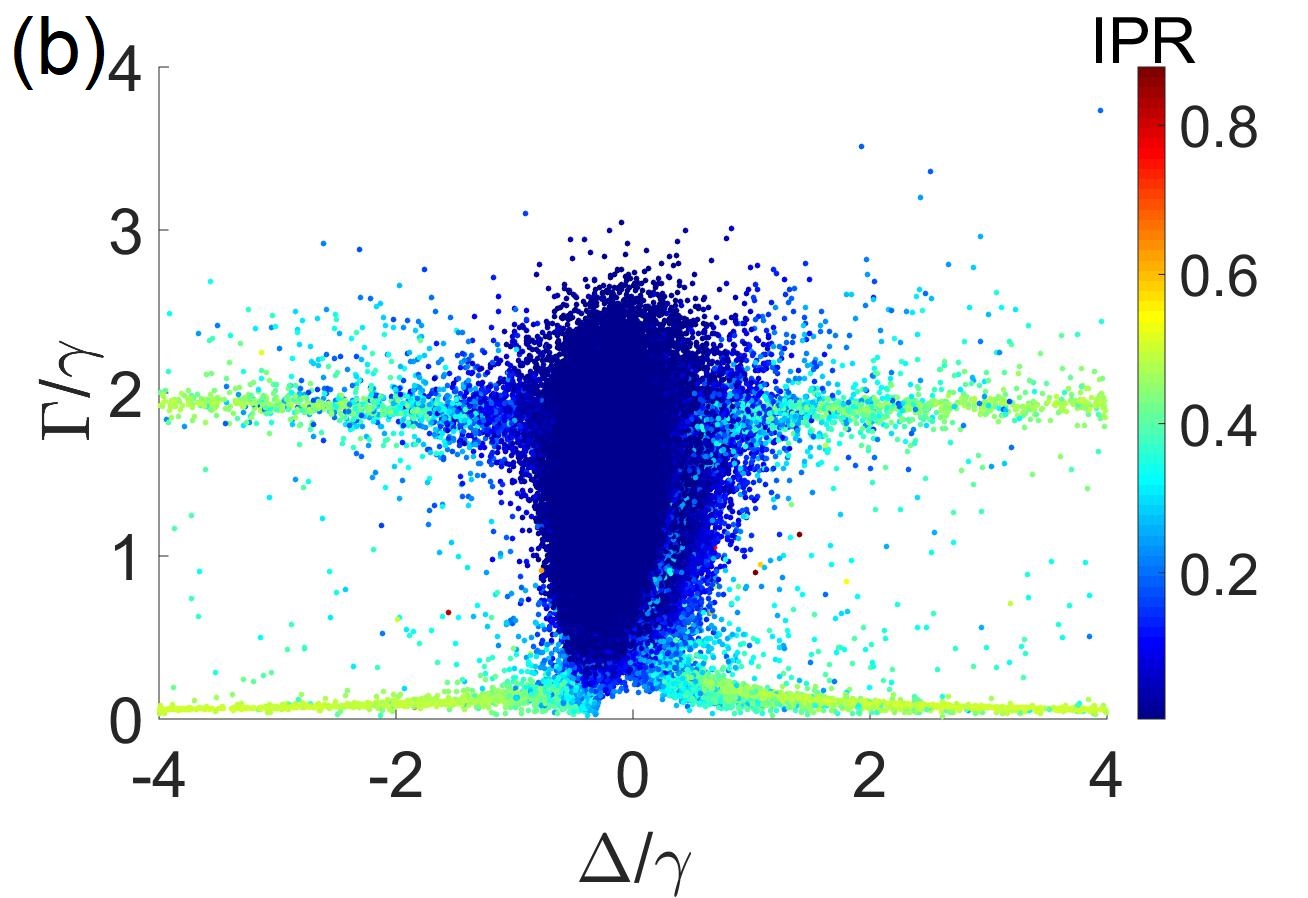}\label{eigenmodecor0min}
	}
	\hspace{0.01in}
	\subfloat{
		\includegraphics[width=0.25\linewidth]{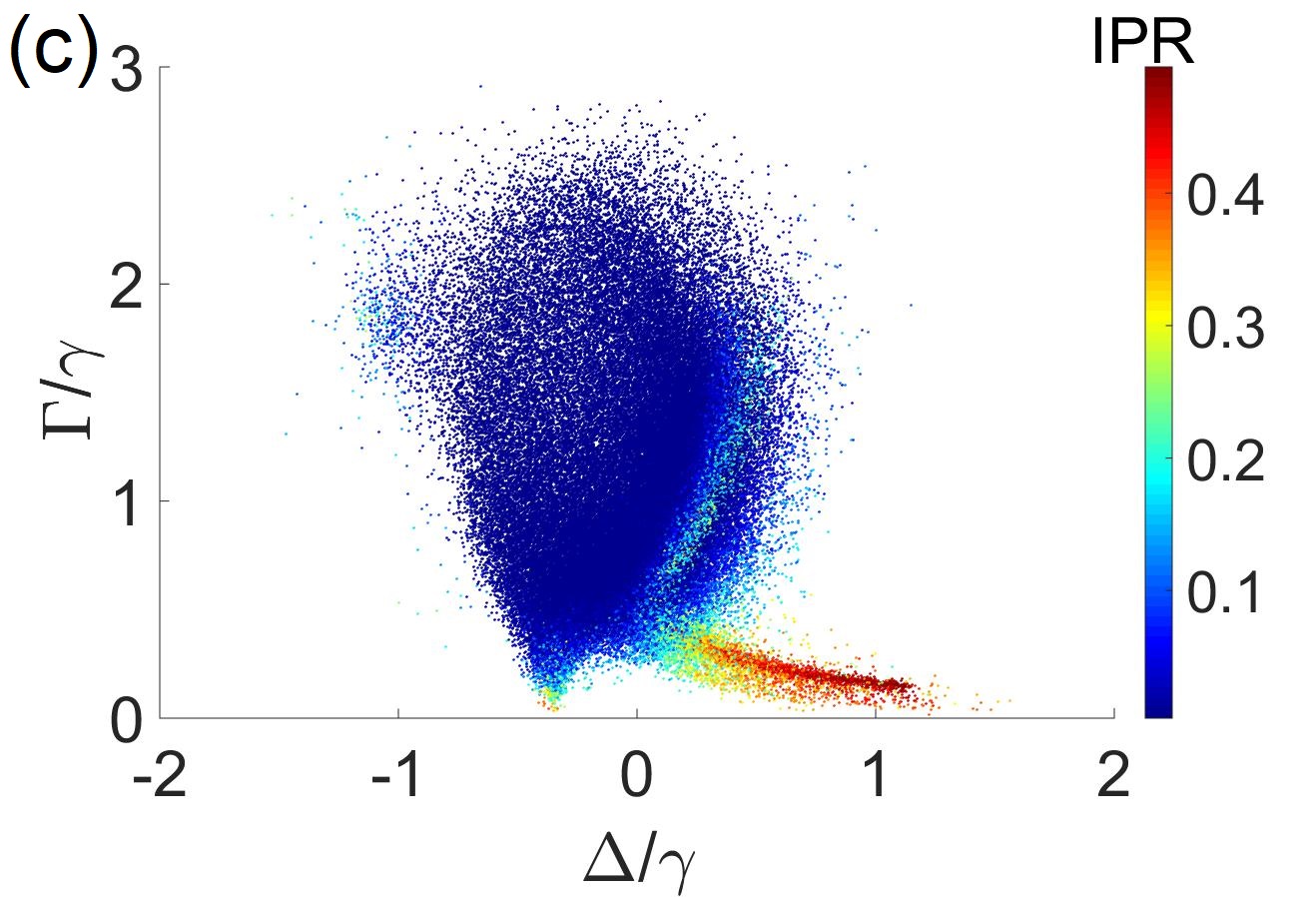}\label{eigenmodecor02}
	}
	\subfloat{
		\includegraphics[width=0.25\linewidth]{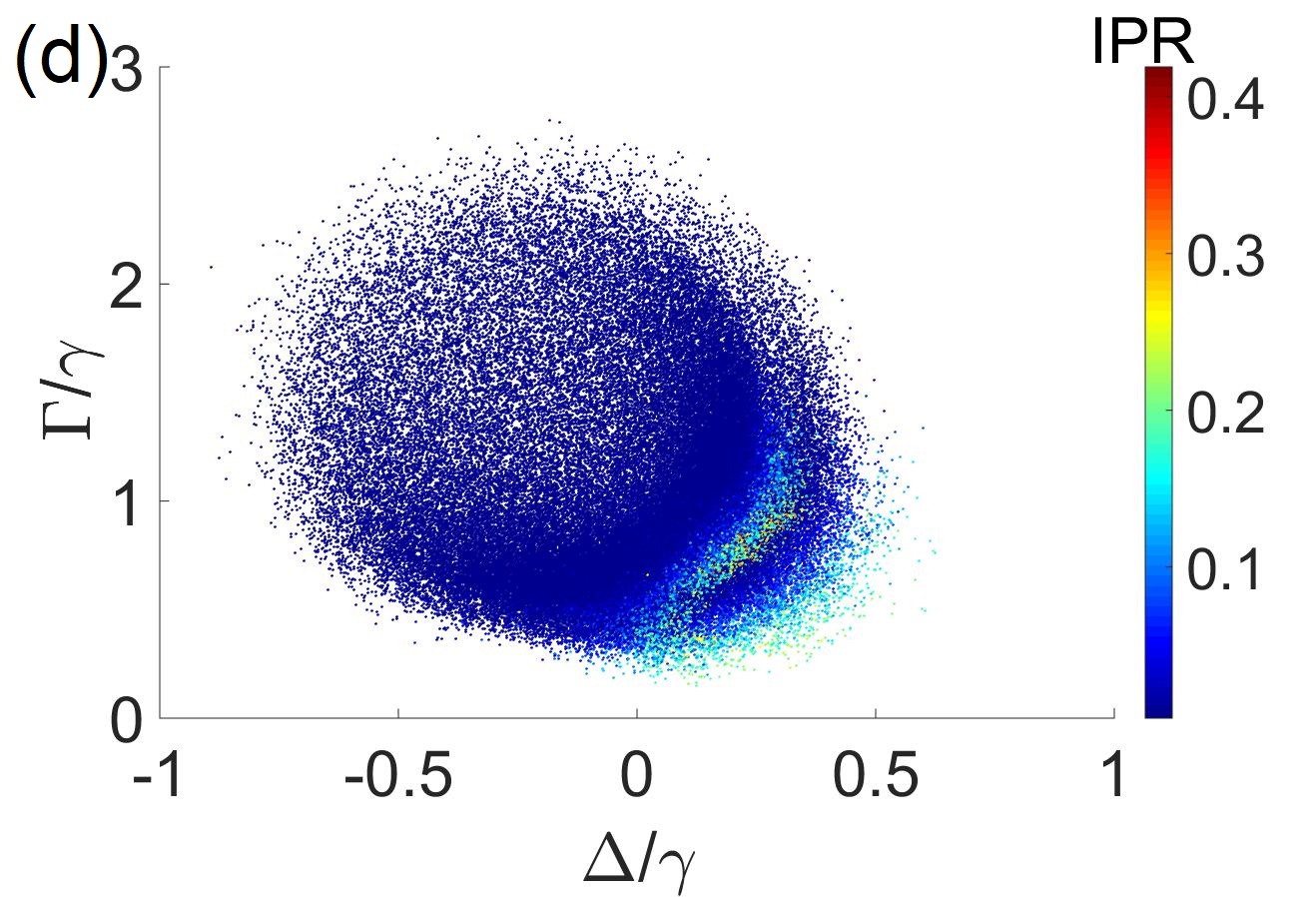}\label{eigenmodecor04}
	}
	\hspace{0.01in}
	\subfloat{
		\includegraphics[width=0.25\linewidth]{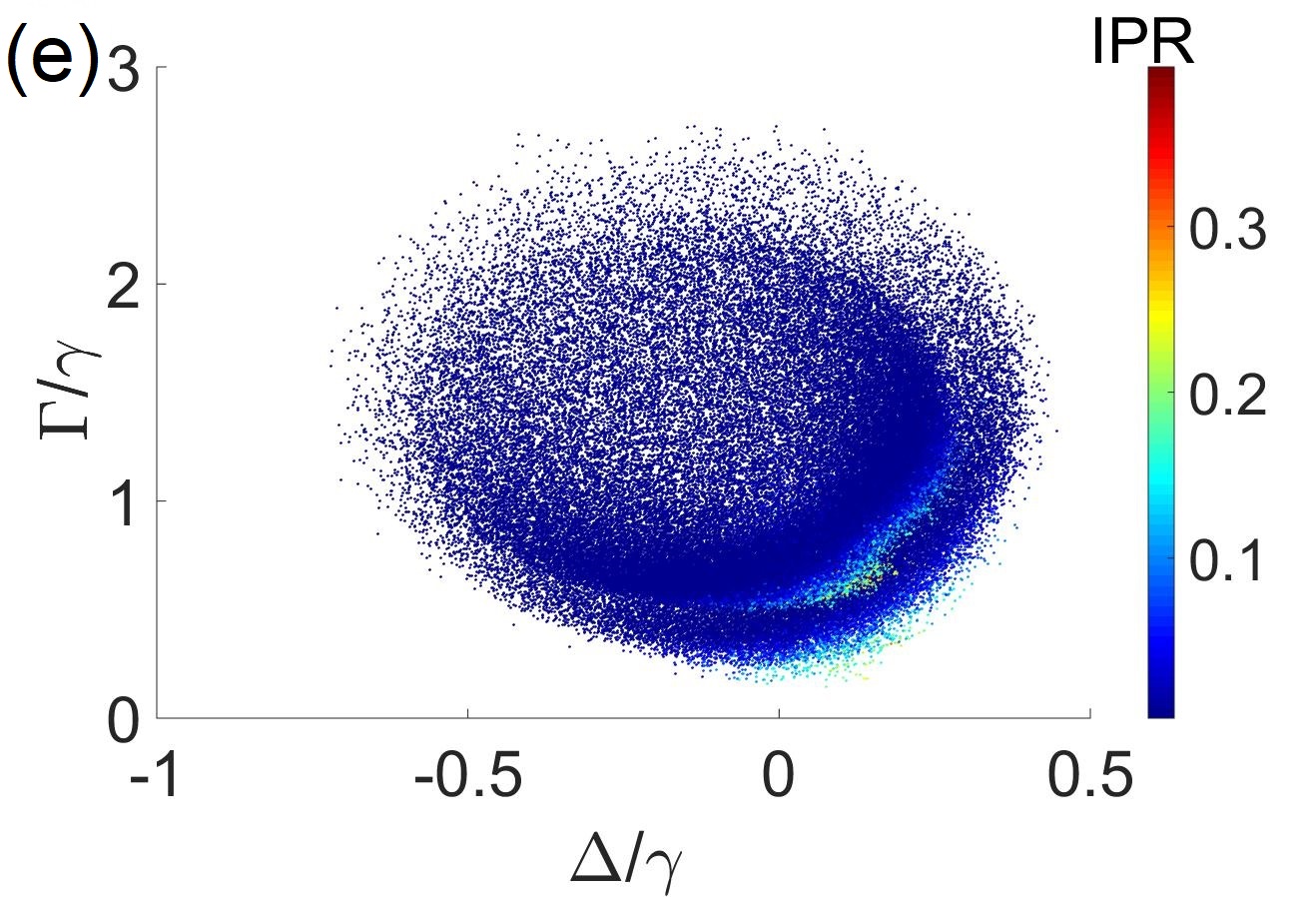}\label{eigenmodecor06}
	}
	\hspace{0.01in}
	\subfloat{
		\includegraphics[width=0.25\linewidth]{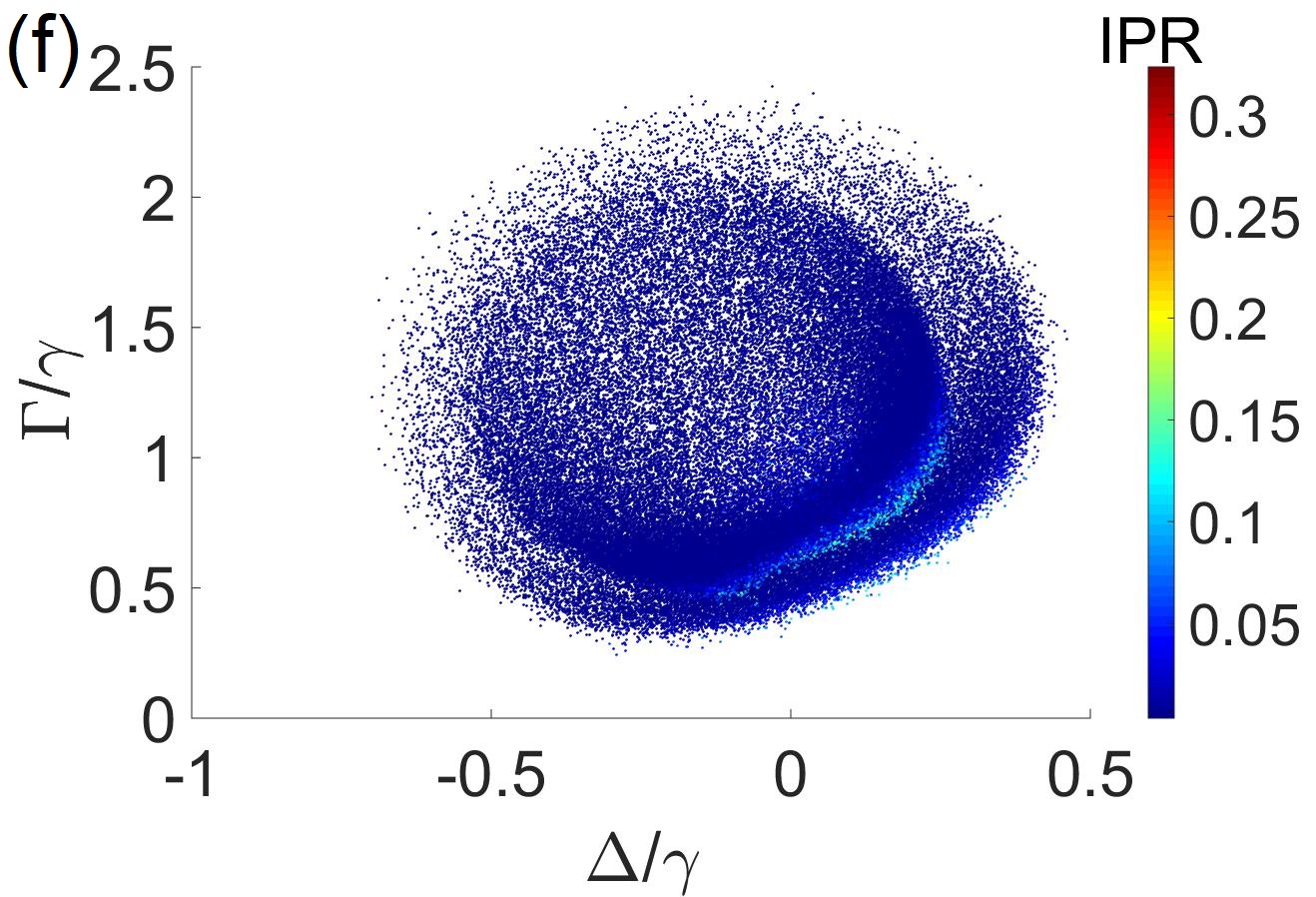}\label{eigenmodecor08}
	}
	\hspace{0.01in}
	\subfloat{
		\includegraphics[width=0.25\linewidth]{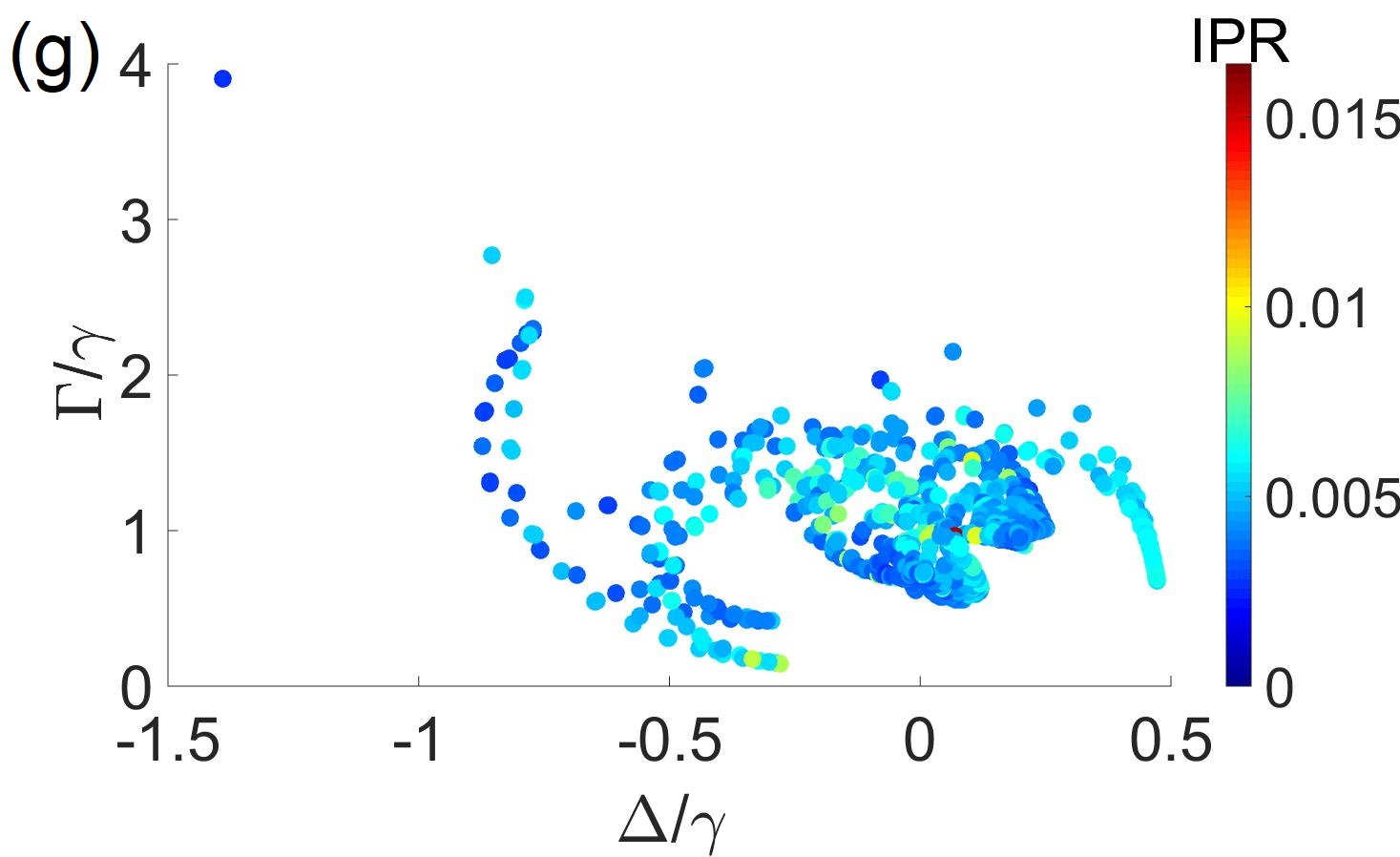}\label{eigenmodeorder}
	}
	\caption{Eigenstate distribution of disordered lattices with different correlation lengths for 10000 realizations. (a) $L_c=0$ (uncorrelated case). (b) An enlarged figure of eigenstates with small detunings for the case of $L_c=0$. (c) $L_c=0.2a$. (d) $L_c=0.4a$. (e)$L_c=0.6a$. (f) $L_c=0.8a$. (g) Ordered lattice. Here $\Delta$ and $\Gamma$ denote the detuning and decay rate of an eigenstate.}\label{eigenstate}
\end{figure}
To further understand the role of positional correlations, we implement a numerical analysis of the eigenstates of the disordered system. The eigenstate can be obtained from the following coupled dipole equations without any incident fields
\begin{equation}\label{coupled-dipole_noinc}
\frac{\omega^2}{c^2}\sum_{i=1,i\neq j}^{N}\mathbf{G}_0(\omega,\mathbf{r}_j,\mathbf{r}_i)\mathbf{d}_i(\omega)=\alpha(\omega)^{-1}\mathbf{d}_j(\omega).
\end{equation}
By expressing the above equations into a matrix notation, namely, using the interaction matrix $\tilde{\mathbf{G}}(\omega)_{ij}=\mathbf{G}_0(\omega,\mathbf{r}_j,\mathbf{r}_i)$ and dipole moment matrix $\tilde{\mathbf{d}}=[\mathbf{d}_1 \mathbf{d}_2 ... \mathbf{d}_N]$, we can obtain the eigen-frequency equation from Eq.(\ref{coupled-dipole_noinc}) in the following form:
\begin{equation}\label{coupled-dipole_eigen}
-\frac{3\pi\gamma}{\omega/c}\tilde{\mathbf{G}}(\omega)\cdot\tilde{\mathbf{d}}(\omega)=(\omega-\omega_0+i\gamma/2)\tilde{\mathbf{d}}(\omega).
\end{equation}
This equation specifies a set of solutions for $\omega$, which are related to the eigenvalues of the Green's function matrix. Specifically, a complex frequency $\tilde{\omega}_n=\omega_n-i\Gamma_n/2$ satisfying above equation corresponds to an eigenstate of the system \cite{wangPRA2018b}. The corresponding eigenvector $\mathbf{e}_n$ can also solved simultaneously. From the eigenvectors, it is possible to compute the inverse participation ratio as
\begin{equation}\label{IPR}
\mathrm{IPR}_n=\frac{\sum_{i=1}^{N}|\mathbf{e}_n(\mathbf{r}_i)|^4}{[\sum_{i=1}^{N}|\mathbf{e}_n(\mathbf{r}_i)|^2]^2}.
\end{equation}
The IPR describes the spatial content of an eigenstate. For instance, for an IPR approaches $1/M$, where $M$ is an integer, the corresponding eigenstate involves the excitation of $M$ atoms \cite{wangPRA2018b}. 

The calculated eigenstate distributions for $a=\lambda_0$ under different correlation lengths are presented in Fig.\ref{eigenstate}. We can recognize that with the increase of correlation lengths, the distribution of eigenstates narrows down to the vicinity of the single atom point ($\Delta=0$ and $\Gamma=\gamma$), namely, both the frequencies and decay rates of the eigenstates approach the behavior of a single atom. This is because the dipole-dipole interactions, which are significant for small interatomic distances, are reduced due to the positional correlations. It is found that for the fully disordered system and the $L_c=0.2a$ case, there are a substantial proportion of eigenstates are highly localized with very small decay rates, which also result from strong dipole-dipole interactions \cite{schilderPRA2016,schilderPRA2017}. Moreover, along with the evolution from a fully random distribution to short-range ordered distributions and then to a long-range ordered distribution, the largest IPR of the eigenstates becomes smaller and smaller, and in the fully ordered case, most of the eigenstates turn to be extended over the entire system.


\begin{figure}[htbp]
	\centering
\subfloat{
	\includegraphics[width=0.42\linewidth]{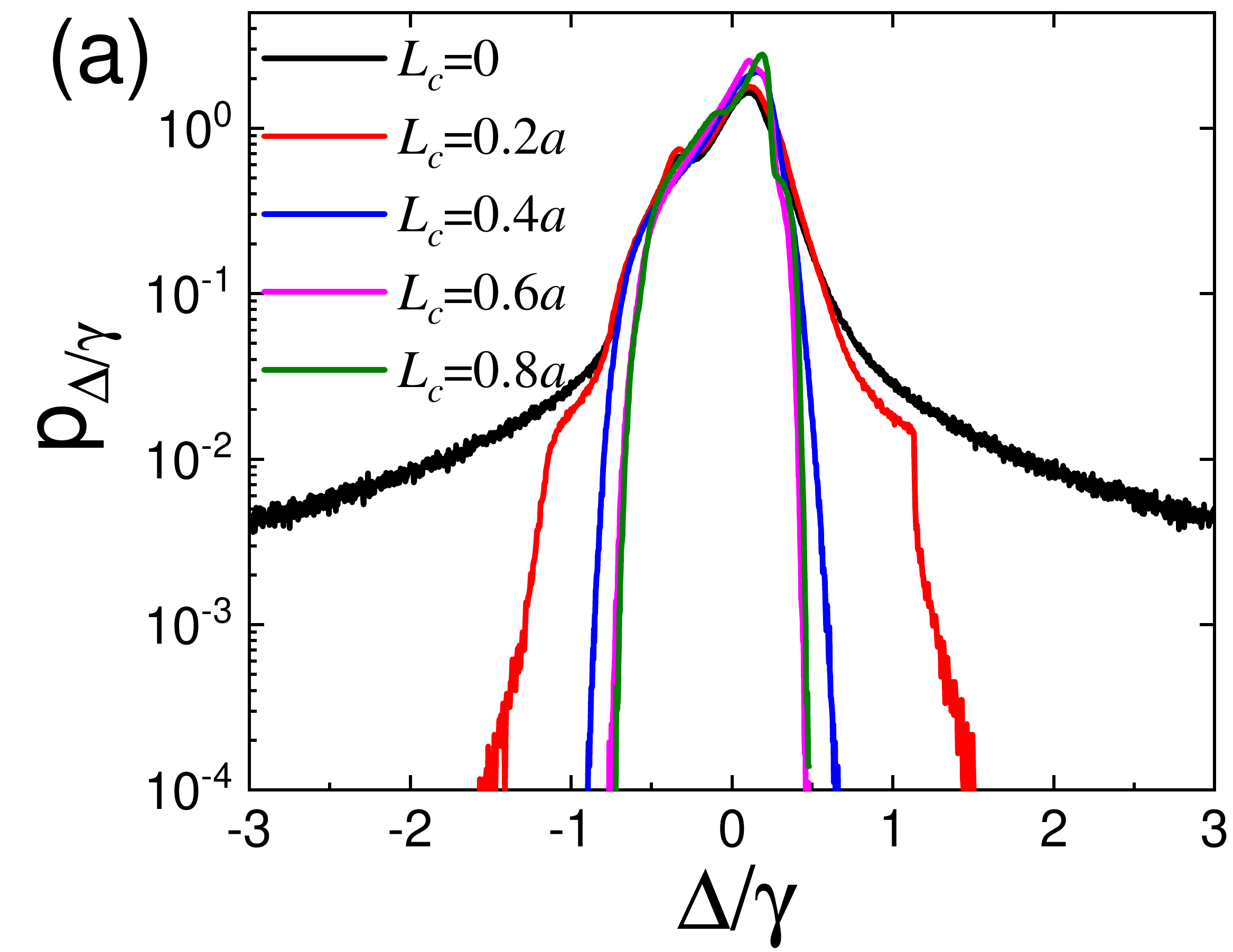}\label{frequencyPDFratio1}
}
	\hspace{0.01in}
\subfloat{
	\includegraphics[width=0.42\linewidth]{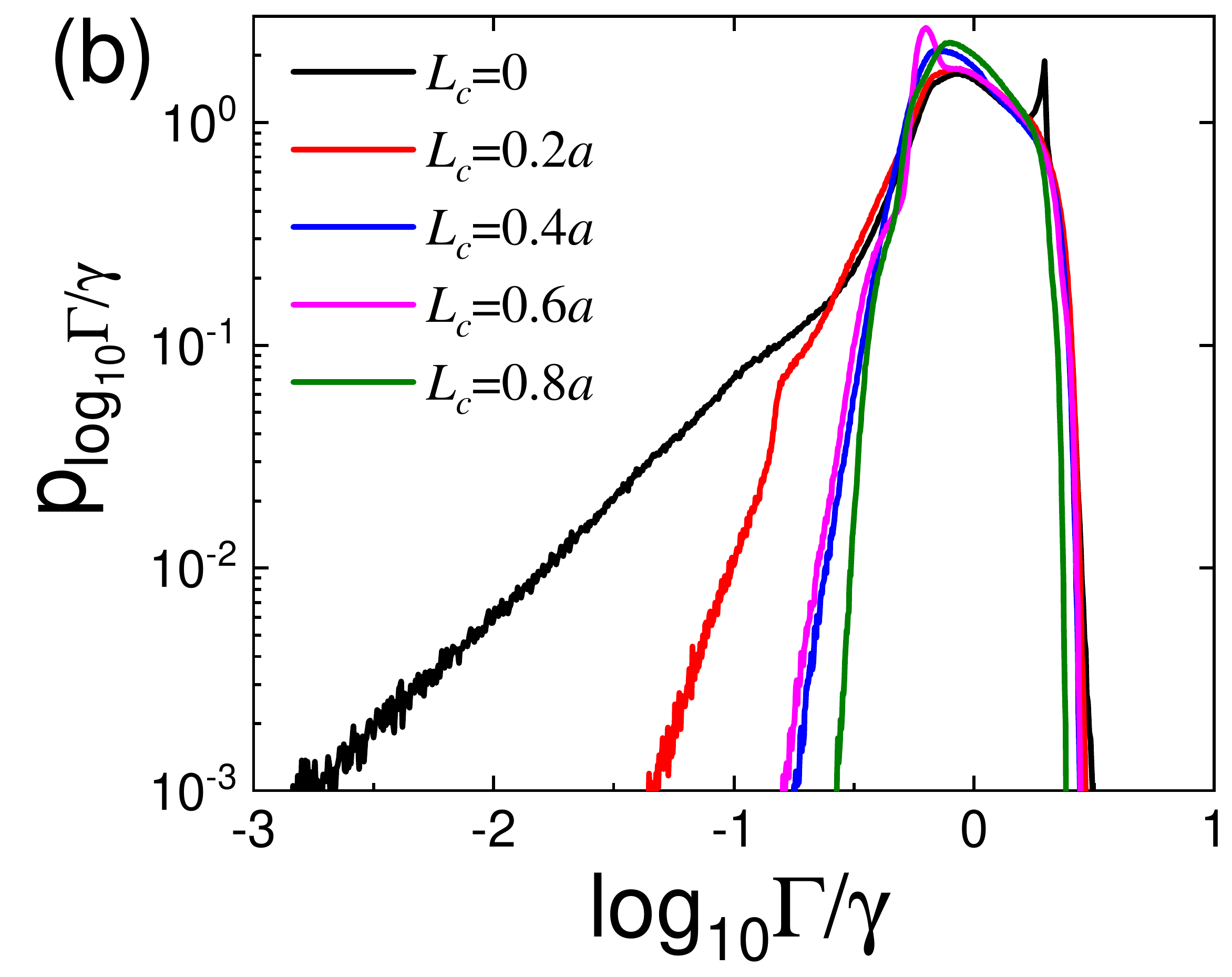}\label{decayratestatisticsratio1}
}
	\caption{Statistical distribution of (a) eigenstate frequencies and (b) decay rates for 2D atomic arrays with different correlation lengths under $a/\lambda_0=1$.}\label{statisticstatio1}
\end{figure}

Above results indicate that the statistical distribution of resonance frequencies and decay rates of the eigenstates is an important signal of the dipole-dipole interactions and the short-range order. To establish a more quantitative understanding, in Figs.\ref{frequencyPDFratio1} and \ref{decayratestatisticsratio1} we show the statistical distribution of resonance frequencies and decay rates under different correlation lengths respectively. This statistics is obtained by performing 10000 disordered realizations for each correlation length. The distribution of the resonance frequencies in a fully random ensemble shows a long tail at large detunings, which can even extend to  $|\Delta/\gamma|>100$, and to be more illustrative, here only the range $|\Delta/\gamma|<3$ is selected. By increasing the correlation length, it can be observed that the eigenfrequency distribution narrows down substantially. In addition, the peaks of the probability density functions of all cases are near the single atom resonance but all slightly blue-shifted, in contrast to the frequency positions of the transmission valleys (or OD peaks) that are inclined to redshift along with the increase of correlation lengths. This is because, although more eigenstates exist at $\Delta>0$, they are less populated by a plane wave incidence due to lower decay rates and thus weaker coupling to free-space radiation, as can be inferred from Fig.\ref{eigenstate}. 

In terms of the decay rate distribution in Fig.\ref{decayratestatisticsratio1}, it can be clearly observed that the smaller the correlation length, the more eigenstates with small decay rates emerge, which constitute the long tails in the statistical distribution.  Note the figure is plotted against $\log_{10}\Gamma/\gamma$ to show the highly nonradiative eigenstates with $\Gamma\rightarrow0$. This phenomenon can provide an explanation for the relatively high transmission of the atomic clouds with small correlation lengths, that is, these nonradiative eigenstates can couple to the vacuum very weakly and result in much smaller scattering and thus total extinction. In addition, it can be noted that for the fully disordered case, there is a peak of $\Gamma/\gamma=2$, which stands for a large proportion of superradiant states formed between a pair of atoms, strongly influenced by the dipole-dipole interaction between atoms which has the $r^{-3}$ behavior in the small atomic distance, while no peaks around $\Gamma/\gamma=2$ are observed in the correlated cases \cite{grossPhysRep1982,fengPRA2013,fengPRA2014}.

\section{Larger atom number densities}
\begin{figure}[htbp]
	\centering
	\subfloat{
		\includegraphics[width=0.3\linewidth]{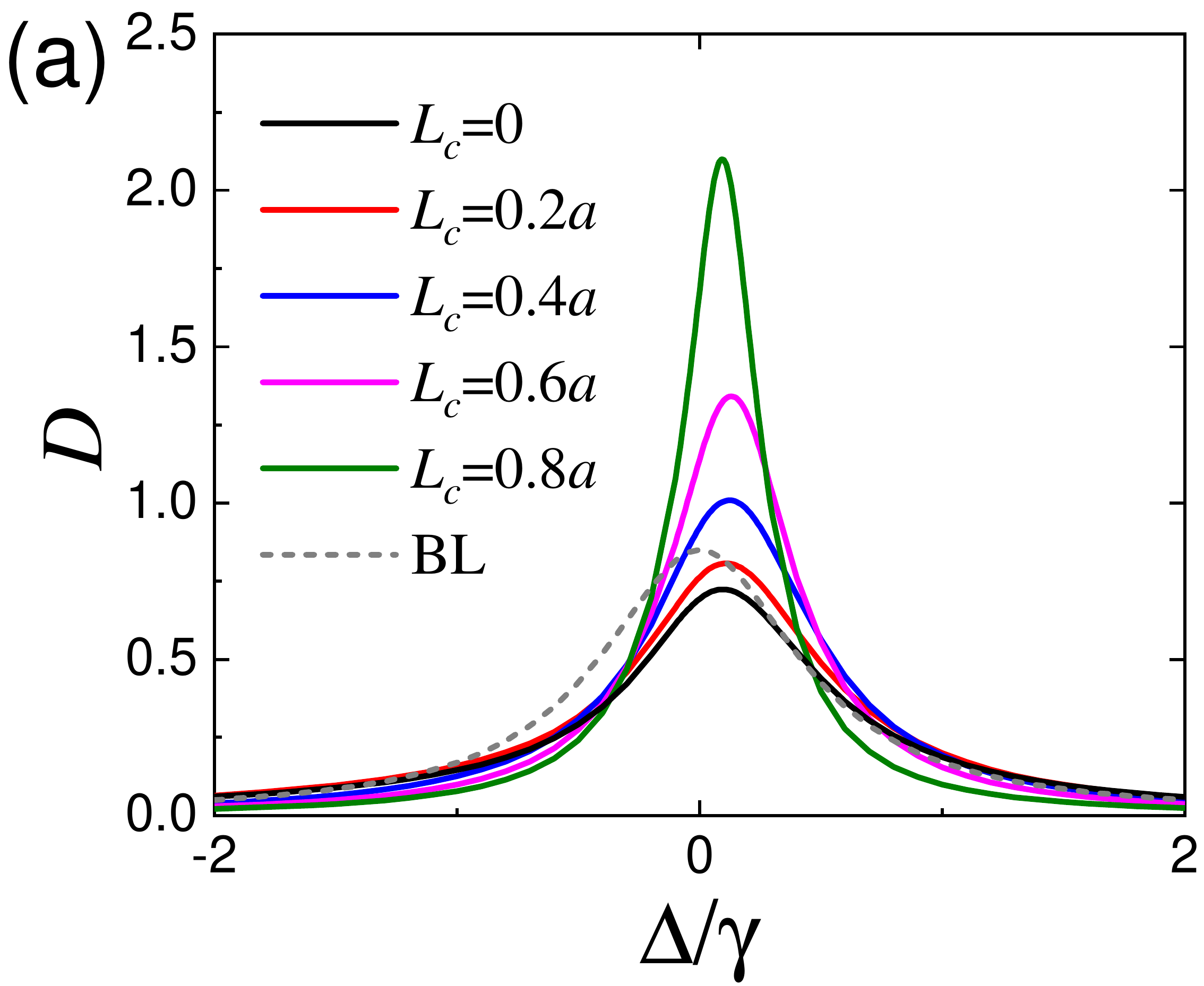}\label{OD_ratio075}
	}
	\hspace{0.01in}
	\subfloat{
		\includegraphics[width=0.3\linewidth]{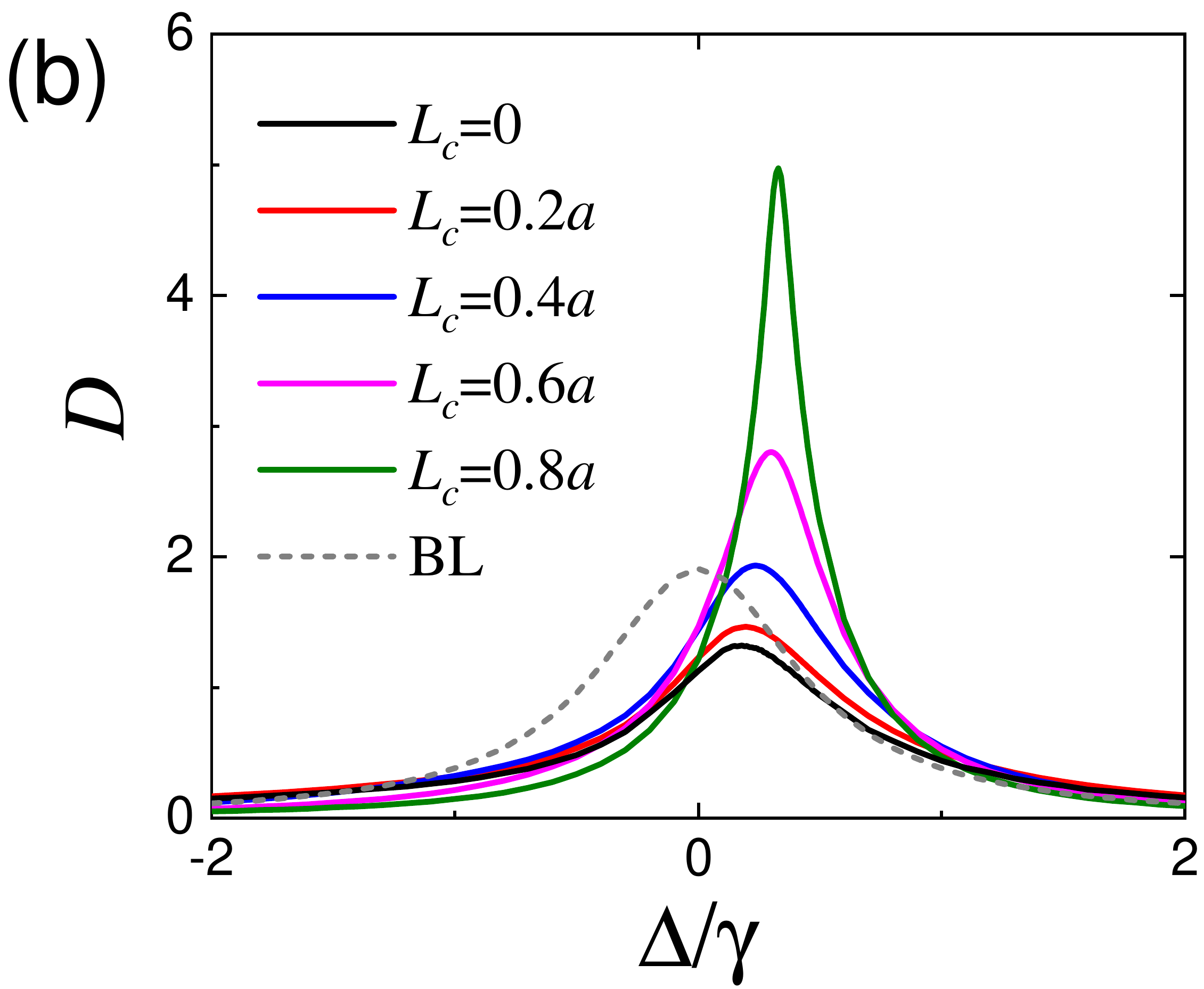}\label{OD_ratio05}
	}
	\hspace{0.01in}
	\subfloat{
	\includegraphics[width=0.3\linewidth]{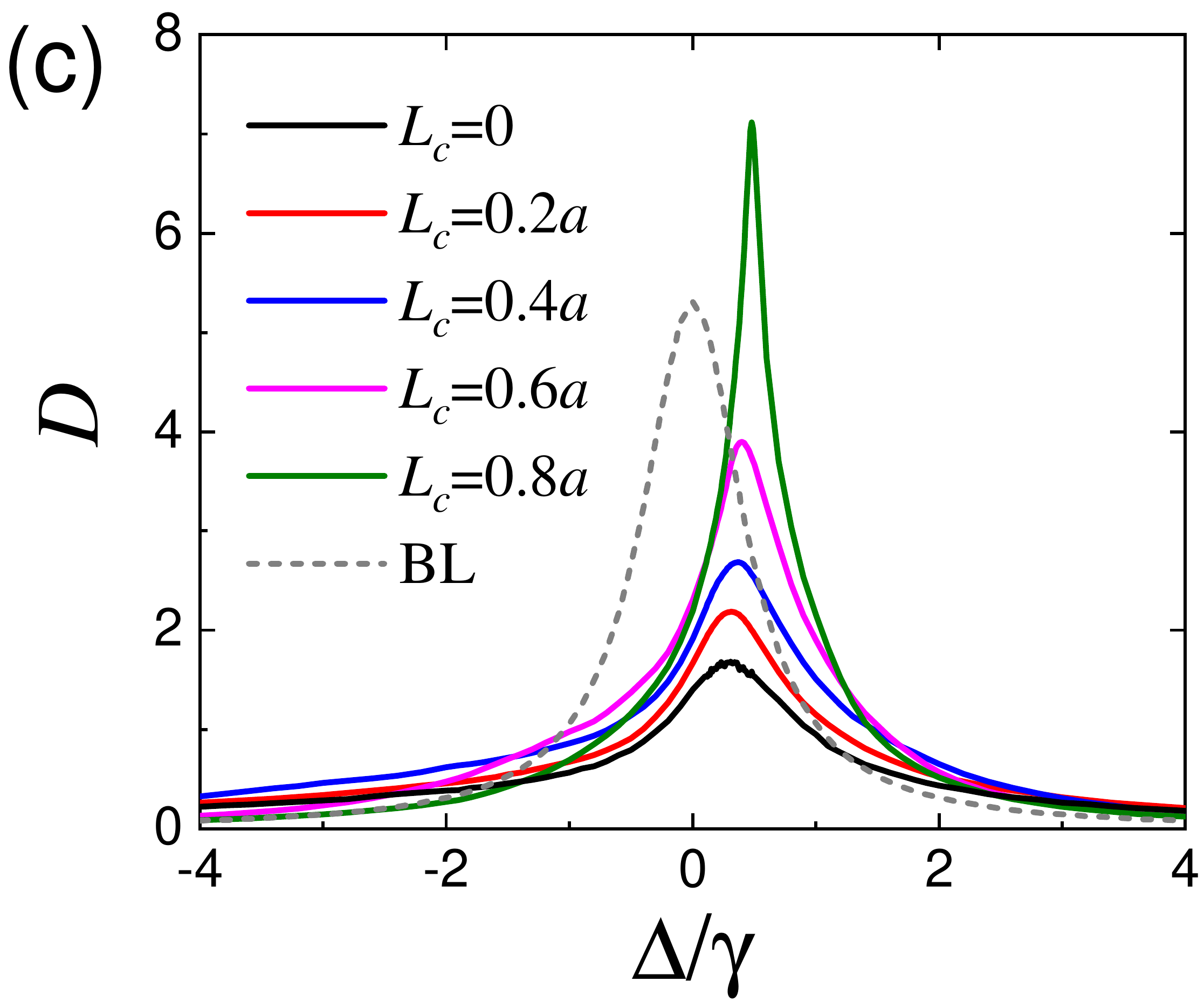}\label{OD_ratio03}
}

	\caption{Optical depth spectra for 2D atomic arrays with different atomic number densities and correlation lengths.  The number densities are represented by (a) $a=0.75\lambda_0$, (b) $a=0.5\lambda_0$ and(c) $a=0.3\lambda_0$, respectively.}\label{ODdensity}
\end{figure}

After establishing a basic understanding on the role of positional correlations, especially the effect on dipole-dipole interactions, we continue to the cases with higher atomic number densities, in which dipole-dipole interactions become more significant. In Fig.\ref{ODdensity}, the optical depth spectra for 2D atomic arrays with different atomic number densities and correlation lengths are shown, in which the number densities are progressively increased and represented through lattice constant $a$ as $a=0.75\lambda_0$,  $a=0.5\lambda_0$ and $a=0.3\lambda_0$, respectively.

\begin{figure}[htbp]
	\centering
	\subfloat{
		\includegraphics[width=0.42\linewidth]{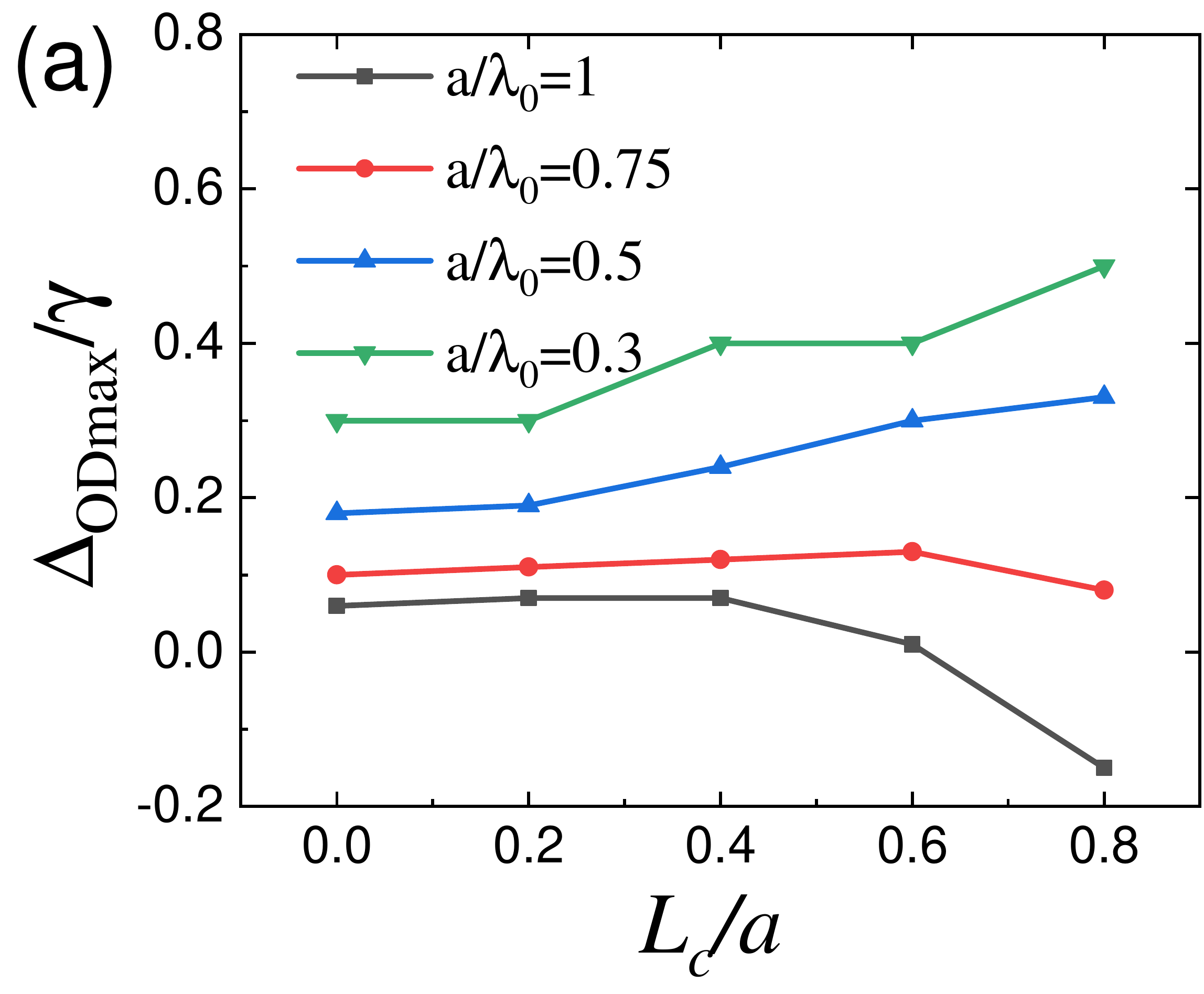}\label{shift_data}
	}
	\hspace{0.01in}
	\subfloat{
		\includegraphics[width=0.42\linewidth]{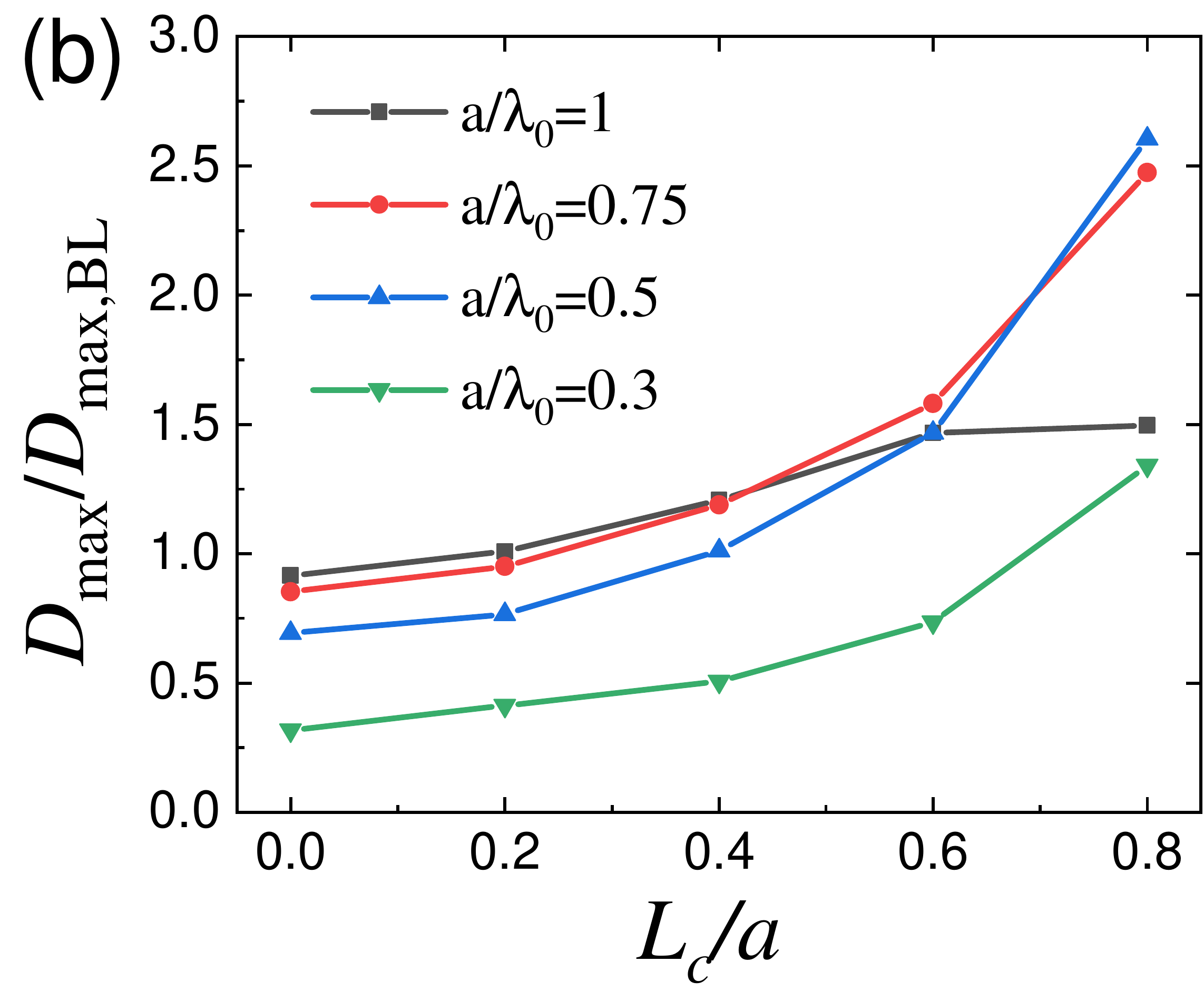}\label{ODmaxratio}
	}
	\caption{(a) The frequency shift of OD peaks and (b) the ratio between the OD peaks and BL predicted values as a function of the correlation length at different number densities.}\label{ODshiftratio}
\end{figure}

It is noted that the behavior of the $a=0.75\lambda_0$ case is similar with that of the $a=\lambda_0$ case. In particular, the peak value of the optical depth spectra of disordered atomic samples grows while the spectral position of the optical depth peak slightly redshifts with the increase of the correlation length. However, in the cases of higher densities (Figs.\ref{OD_ratio05} and \ref{OD_ratio03}), the spectral position of the OD peak turns to be progressively blueshifted for stronger positional correlations. This phenomenon, which is also summarized in Fig.\ref{shift_data} plotting the frequency shift of OD peaks $\Delta_\mathrm{ODmax}/\gamma$, comes from the intricate competition between significant dipole-dipole interactions at high number densities and positional correlations. That is, the dipole-dipole interactions tend to blueshift the spectra while positional correlations are inclined to make the spectra redshifted. At high atom number densities, the dipole-dipole interactions play a more significantly role than positional correlations. This observation is also consistent with that of Chomaz et al's \cite{chomazNJP2012}, which implied that the blueshift is indeed a many-body phenomenon. This feature can be further summarized in Fig.\ref{ODmaxratio}, which provides the maximum of OD, $D_\mathrm{max}$, normalized by the BL prediction $D_\mathrm{max,BL}=6\pi n_0/k^2$, as a function of the correlation length at different atom number densities. It is demonstrated that positional correlations can lead to an enhancement to $D_\mathrm{max}$ while dipole-dipole interactions in turn reduce this enhancement, leading to the lowest enhancement in the case of $a=0.3\lambda_0$.

\begin{figure}[htbp]
	\centering
	\subfloat{
		\includegraphics[width=0.3\linewidth]{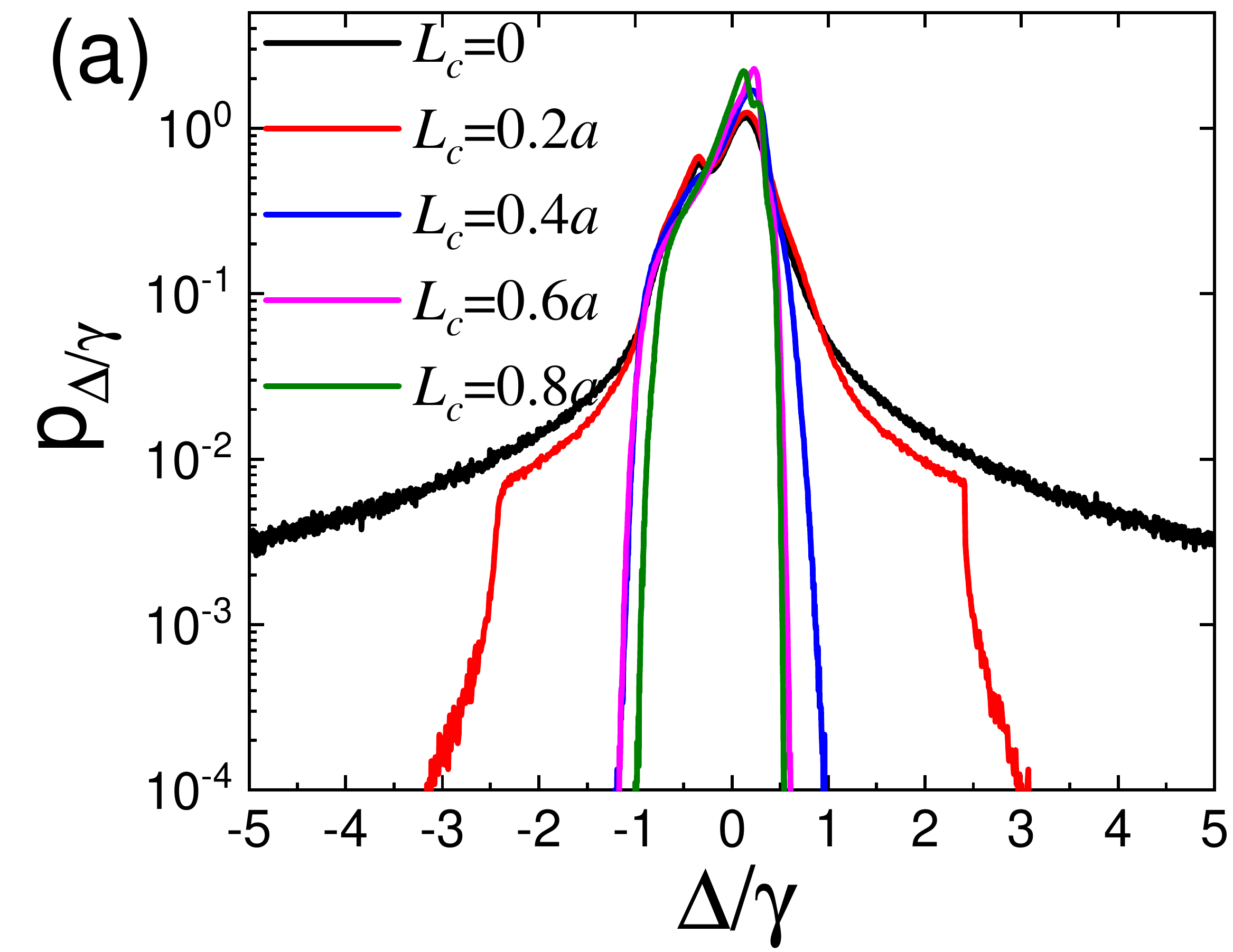}\label{frequencyPDFratio075}
	}
	\hspace{0.01in}
	\subfloat{
		\includegraphics[width=0.3\linewidth]{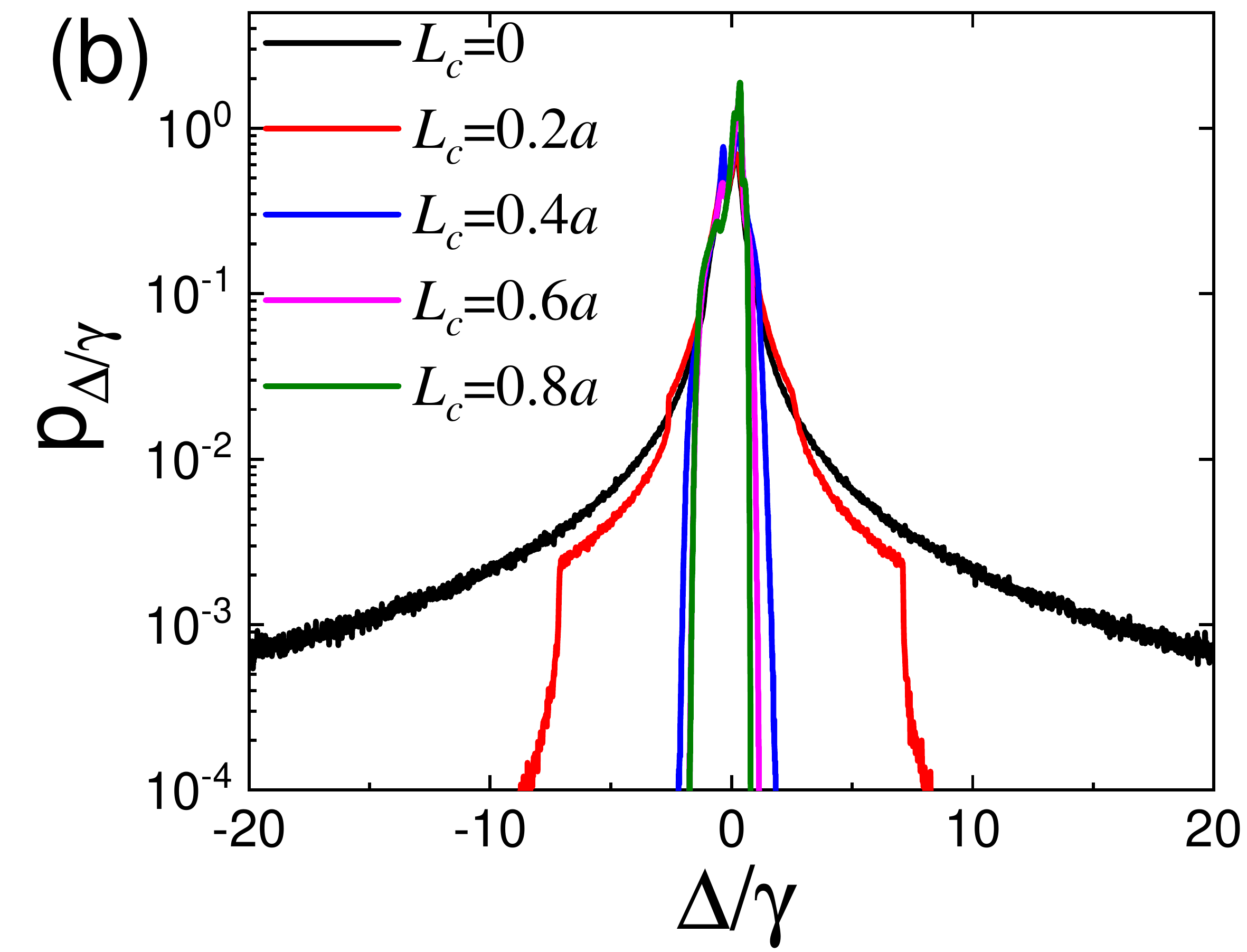}\label{frequencyPDFratio05}
	}
	\hspace{0.01in}
	\subfloat{
		\includegraphics[width=0.3\linewidth]{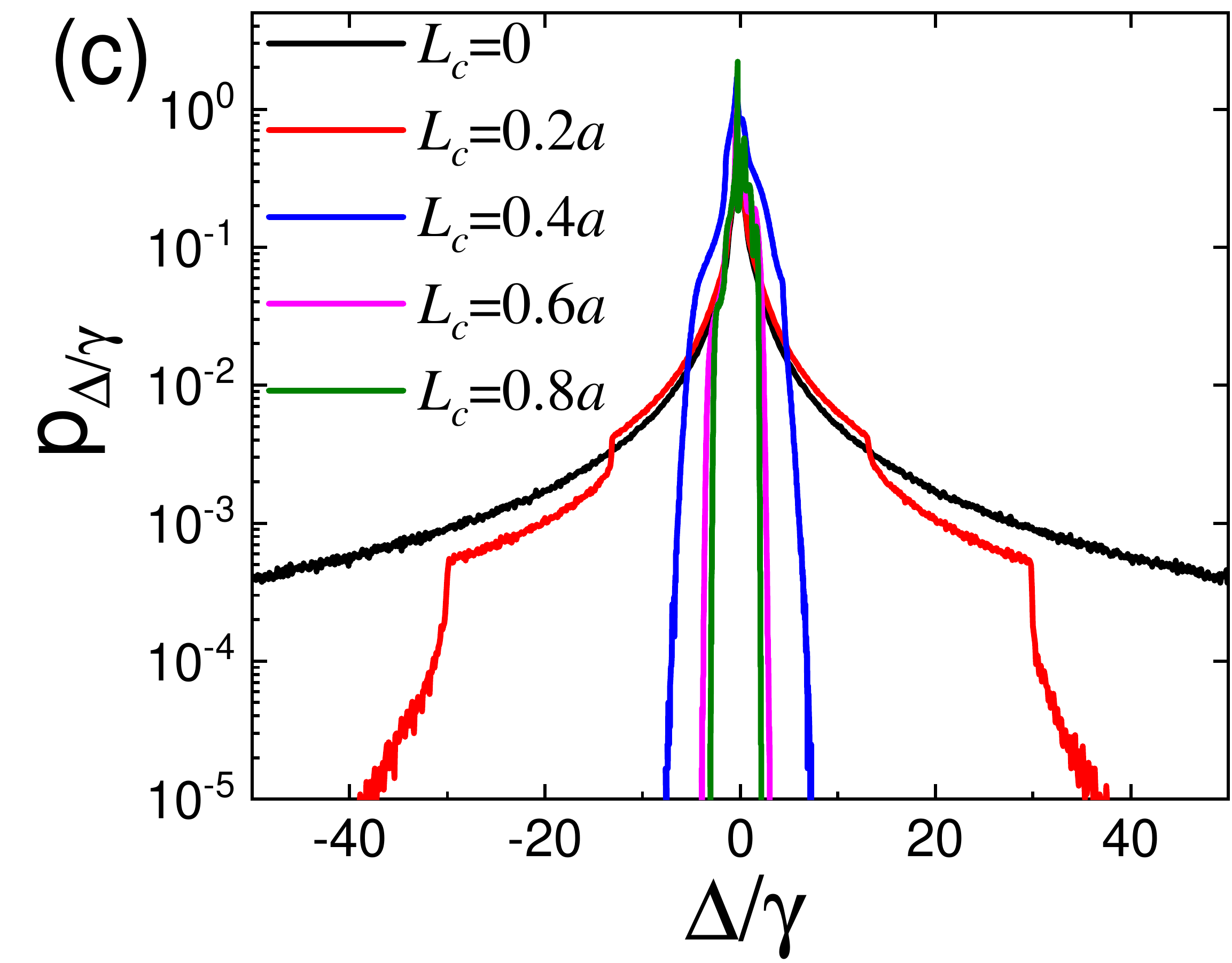}\label{frequencyPDFratio03}
	}
	\caption{Eigenstate frequency statistics for 2D atomic arrays with different atomic number densities and correlation lengths. The number densities are represented by (a) $a=0.75\lambda_0$, (b) $a=0.5\lambda_0$ and(c) $a=0.3\lambda_0$, respectively.}\label{frequencystatisticsdensity}
\end{figure}

\begin{figure}[htbp]
	\centering
	\subfloat{
		\includegraphics[width=0.3\linewidth]{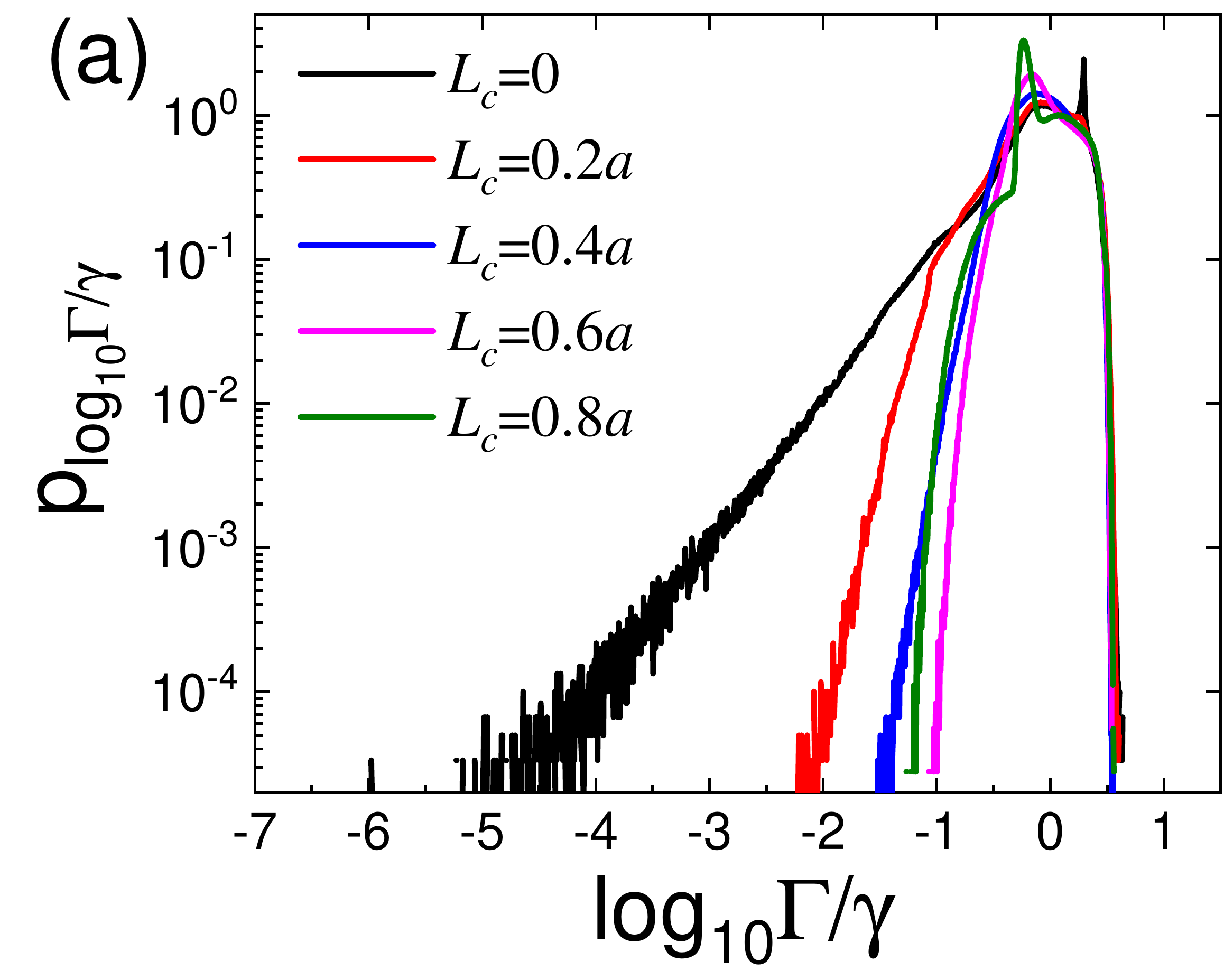}\label{decayratestatisticsratio075}
	}
	\hspace{0.01in}
	\subfloat{
		\includegraphics[width=0.3\linewidth]{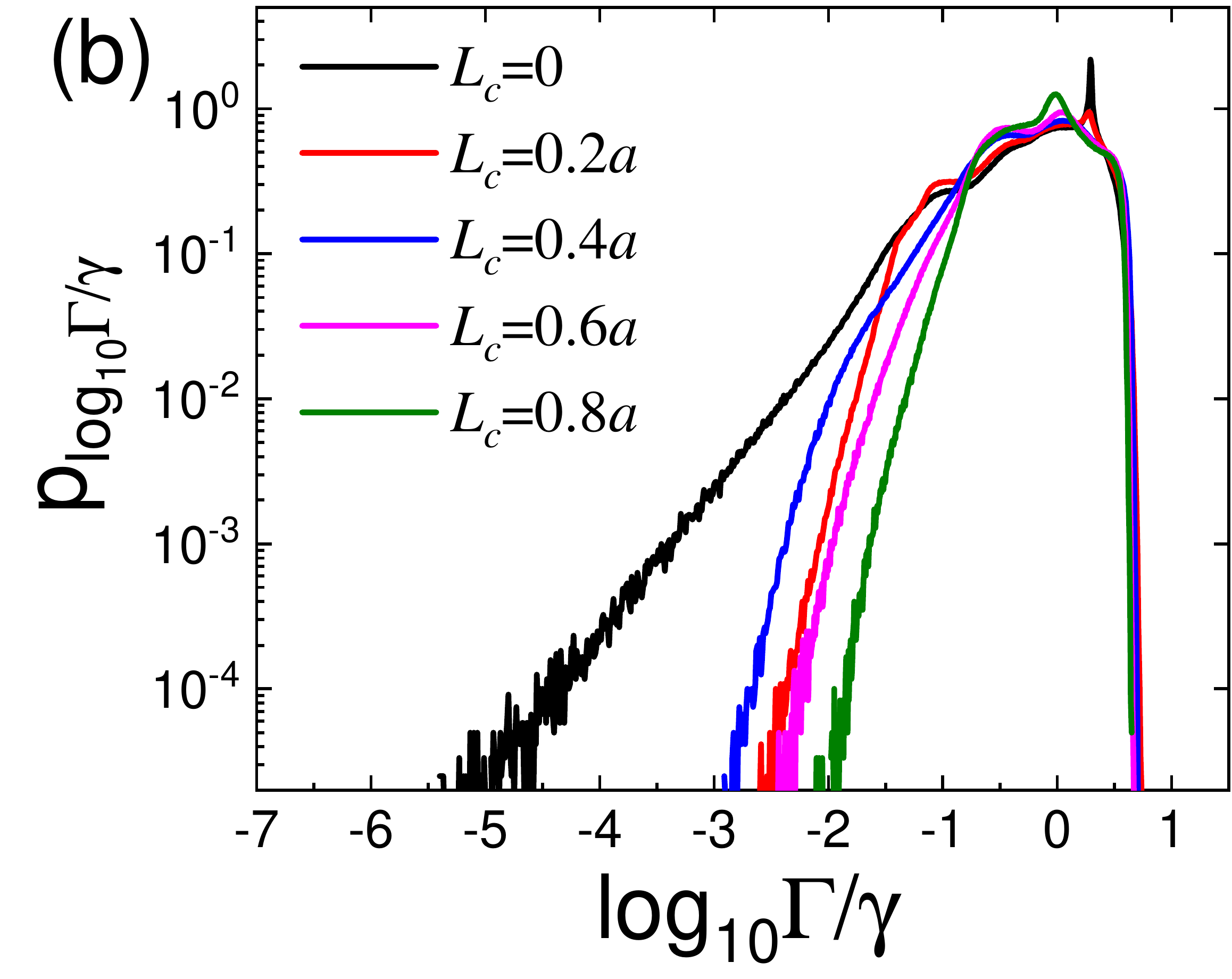}\label{decayratestatisticsratio05}
	}
	\hspace{0.01in}
	\subfloat{
	\includegraphics[width=0.3\linewidth]{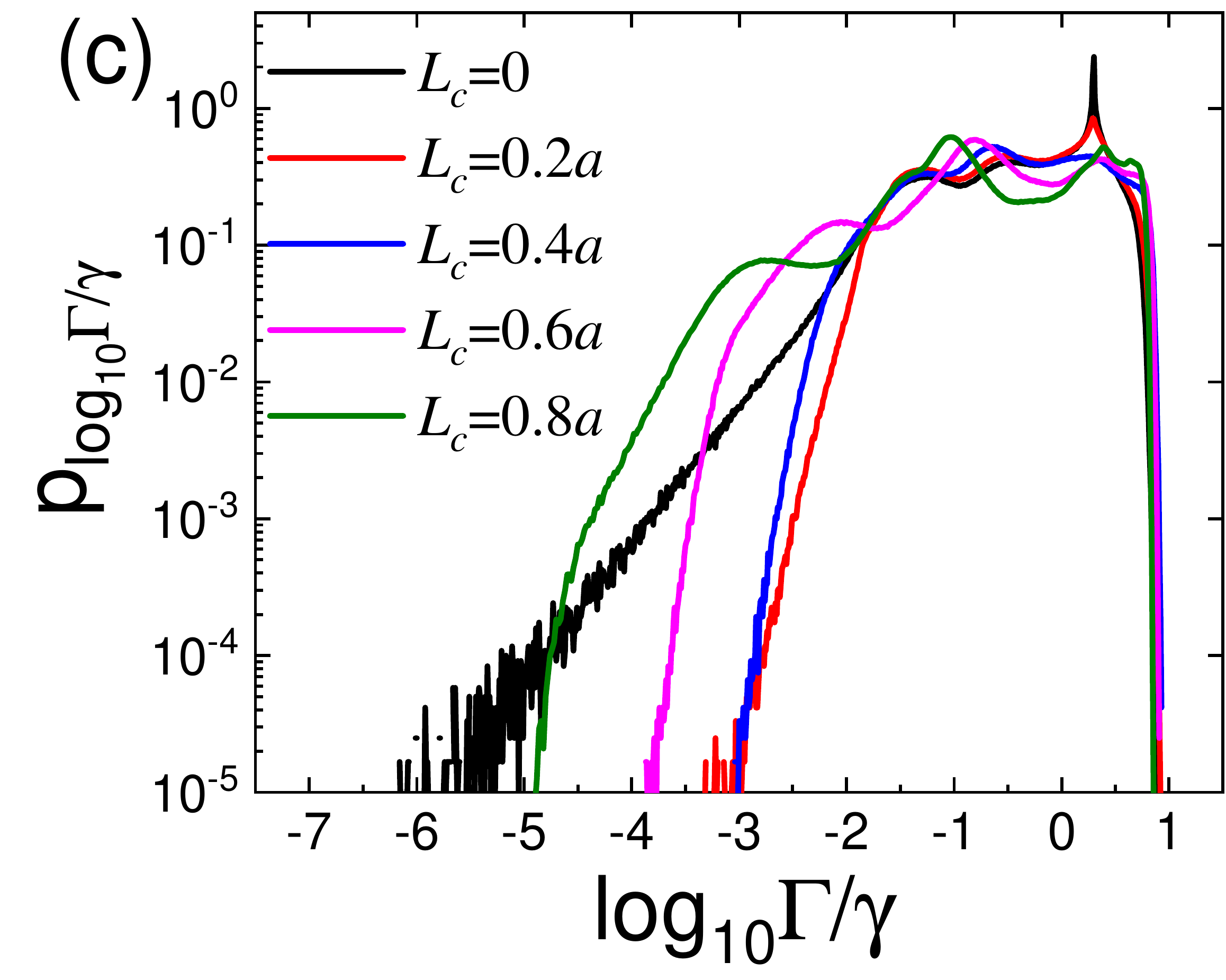}\label{decayratestatisticsratio03}
	}
	\caption{Decay rate statistics for 2D atomic arrays with different atomic number densities and correlation lengths. The number densities are represented by (a) $a=0.75\lambda_0$, (b) $a=0.5\lambda_0$ and (c) $a=0.3\lambda_0$, respectively.}\label{decayratestatisticsdensity}
\end{figure}

In Figs.\ref{frequencystatisticsdensity} and \ref{decayratestatisticsdensity} we continue to show the statistical distribution of resonance frequencies and decay rates for different atom number densities, respectively. It is revealed in Fig.\ref{frequencystatisticsdensity} that a higher number density leads to a more broadened eigenfrequency distribution, and the long tails in the $L_c=0$ cases also become more extended to large detunings. Note the long tails of the uncorrelated cases are not fully shown to better identify the statistics of correlated cases. This result is obviously due to the dipole-dipole interactions that has a $r^{-3}$ behavior in the small atomic distance and lead to more significant frequency shifts \cite{fengPRA2014,schilderPRA2016}.

Regarding the decay rate distribution in Fig.\ref{decayratestatisticsdensity} under different atom number densities, more interesting behaviors can be found. In the highest density case (Fig.\ref{decayratestatisticsratio03}), it is observed the dependence of decay rate statistics on the correlation length is almost reversed in comparison with the dilute case in Fig.\ref{decayratestatisticsratio1}. To be more specific, except for the uncorrelated case (in which the behaviors of a set of pairs of atoms dominate), with the increase of correlation length, the decay rate distribution becomes more extended towards the very small decay rate range. In the meanwhile, for intermediate atom number densities (Figs.\ref{decayratestatisticsratio075} and \ref{decayratestatisticsratio05}), the dependence on the correlation length is not monotonous. This phenomenon can be explained by the fact that at high atom number densities where the distances among atoms are small compared to the wavelength (i.e., strong dipole-dipole interactions always persist), the introduction of positional correlations combined with a large number of atoms in the wavelength scale can lead to the formation of long-lived collective modes with small decay rates, which are much less possible for small atom number densities. This is also the condition in which for certain disordered media, by engineering appropriate positional correlations and subwavelength mean distance between scatterers, photonic band gaps can form \cite{Froufe-Perez2016,ricouvierPNAS2019}. Therefore, our results can provide a general picture of the complex interplay between dipole-dipole interactions and positional correlations.

\section{Conclusions}
In summary, we analyze near-resonant light transmission in two-dimensional dense ultracold atomic ensembles with short-range positional correlations to understand the cooperative and collective effects in a random collection of atoms. Based on the coupled-dipole simulations under different atom number densities and correlation lengths, we show that the collective effects are strongly influenced by those positional correlations, manifested as significant shifts and broadening or narrowing of transmission resonance lines. We also analyze the eigenstate distribution of different atomic ensembles, in which the statistics of eigenfrequencies and decay rates is strongly affected by the interplay between positional correlations and dipole-dipole interactions. This work may provide profound implications on collective and cooperative effects in cold atomic ensembles as well as the study of mesoscopic physics concerning light transport in strongly scattering disorder media.

\section*{Funding}
The National Natural Science Foundation of China (51636004, 51906144); Shanghai Key Fundamental Research Grant (18JC1413300); China Postdoctoral Science Foundation (BX20180187, 2019M651493); The Foundation for Innovative Research Groups of the National Natural Science Foundation of China (51521004).

\section*{Disclosures}
The authors declare that there are no conflicts of interest related to this article.

\bibliography{disorderatom}


\end{document}